\documentclass[12pt]{article}
\pdfoutput=1
\usepackage{amsmath,amssymb,graphicx} 
\usepackage[footnotesize]{caption}
\usepackage{epsf}
\usepackage{pstricks}
\usepackage{ctable}
\usepackage[sort&compress,numbers]{natbib}
\usepackage{hyperref}
\usepackage{hypernat}
\usepackage{setspace}
\bibpunct{[}{]}{,}{n}{}{;}

\newcommand{\beq}{\begin{eqnarray}}
\newcommand{\eeq}{\end{eqnarray}}
\newcommand{\Br}{{\rm Br}}

\newcommand\ZZ{\hbox{\zfont Z\kern-.4emZ}}
\font\zfont = cmss10 

\newcommand{\met}{\mbox{${\rm \not\! E}_{\rm T}$}}
\newcommand{\dzero}{D$\varnothing$~}
\newcommand{\dz}{D$\varnothing$}
\def\gappeq{\mathrel{ \rlap{\raise.5ex\hbox{$>$}}
                      {\lower.5ex\hbox{$\sim$}}  } }
\def\lappeq{\mathrel{ \rlap{\raise.5ex\hbox{$<$}}
                      {\lower.5ex\hbox{$\sim$}}  } }
\def\CO {{\cal O}}

\def\CA{{\cal A}}

\def\CO {{\cal O}}

\begin{document}
\begin{titlepage}
\begin{flushright}
{\small YITP-SB-09-39}
\end{flushright}

\vskip.5cm
\begin{center}
{\huge \bf Prompt Decays of General}

\medskip

{\huge \bf Neutralino NLSPs at the Tevatron}

\vskip.1cm
\end{center}
\vskip0.2cm

\begin{center}
{\bf Patrick Meade$^{a,b}$, Matthew Reece$^c$ and David Shih$^{b,d}$}
\end{center}
\vskip 8pt

\begin{center}
{\it $^a$C. N. Yang Institute for Theoretical Physics,\\ Stony Brook University, 
Stony Brook, NY 11794} \\
\vspace*{0.1cm}
{\it $^b$Institute for Advanced Study\\
Princeton, NJ 08540} \\
\vspace*{0.1cm}
{\it $^c$Princeton Center for Theoretical Science\\
Princeton University, Princeton, NJ 08544, USA}\\
\vspace*{0.1cm}
{\it $^d$ Department of Physics and Astronomy, \\
Rutgers, 136 Frelinghuysen Rd, Piscataway, NJ 08854}\\
\vspace*{0.1cm}

\end{center}

\vglue 0.3truecm

\begin{abstract}
\vskip 5pt \noindent  

\singlespacing
Recent theoretical developments have shown that gauge mediation has a much larger parameter space of possible spectra and mixings than previously considered. Motivated by this, we explore the collider phenomenology of gauge mediation models where a general neutralino is the lightest MSSM superpartner (the NLSP), focusing on the potential reach from existing and future Tevatron searches. Promptly decaying general neutralino NLSPs can give rise to final states involving missing energy plus photons, Zs, Ws and/or Higgses. We survey the final states and determine those where the Tevatron should have the most sensitivity. We then estimate the reach of existing Tevatron searches in these final states and discuss new searches (or optimizations of existing ones) that should improve the reach. Finally we comment on the potential for discovery at the LHC.

\end{abstract}

\end{titlepage}

\newpage

\renewcommand{\thefootnote}{(\arabic{footnote})}
\onehalfspacing

\section{Introduction}\label{sec:intro}
\setcounter{equation}{0} \setcounter{footnote}{0}

\subsection{Motivation}

Supersymmetry is a promising solution to the gauge hierarchy problem. Its minimal realization -- the MSSM -- predicts a host of new particles  with masses at the weak scale. These particles are presumably already being produced at the Tevatron, and they will be copiously produced at the LHC. If they are someday observed in one of these experiments, then the questions of how SUSY is broken and how its breaking is mediated to the MSSM will become the central puzzles in particle physics.  

SUSY-breaking in the MSSM introduces $\sim 100$ new  parameters beyond those of the Standard Model.  The majority of these parameters introduce flavor-changing effects that are highly constrained by existing experiments. These imply that the mediation of SUSY-breaking to the MSSM must be essentially flavor blind. This is a puzzle in gravity mediation, where one must suppress flavor-changing effects from in principle arbitrary Planck-suppressed higher-dimension operators. On the other hand, it is an automatic consequence of gauge mediation \citep{Dimopoulos:1981au,Dine:1981za,Dine:1981gu,Nappi:1982hm,AlvarezGaume:1981wy,Dimopoulos:1982gm}. In this framework, the SUSY-breaking sector only communicates with the MSSM via the usual SM gauge interactions. Since these interactions are flavor blind, this guarantees that the parameters of the MSSM (in particular the squark and slepton masses) are diagonal in flavor space.

Models of gauge mediation offer a rich variety of novel and distinctive collider signatures.  Perhaps the most characteristic feature of gauge mediation is the light gravitino $m_{3/2}\ll M_{weak}$.  This  ensures that the effects of gauge mediation dominate over the Planck-suppressed contributions of gravity mediation. It implies that the lightest MSSM superpartner is actually the NLSP, which can always decay to the gravitino plus its SM partner. The gravitino will always escape the detector, leading to significant missing energy. Meanwhile, the SM partner will tend to be central and energetic (assuming that the NLSP decays promptly), since it is typically much lighter than its NLSP parent. Given that pairs of NLSPs are produced in every SUSY event, it is clear from these considerations that the inclusive collider signatures of gauge mediation are largely determined by the nature of the NLSP.

This fact has led to many searches for gauge mediation in the inclusive $\gamma\gamma+\met$ channel, especially after the discovery of anomalous Run I events  at the Tevatron that contained $ee\gamma\gamma+\met$ \citep{Abe:1998up}.  Searching for new physics in the inclusive $\gamma\gamma+\met$ channel was also well motivated from gauge mediation models of the mid 90's \citep{Dine:1993yw,Dine:1994vc,Dine:1995ag}.  In these ``ordinary" or ``minimal" gauge mediation models, there are weakly-coupled, ultraheavy ``messenger" fields $\Phi$, $\tilde\Phi$ transforming in complete multiplets of $SU(5)_{\rm GUT}$ which communicate with the SUSY-breaking sector at tree-level.  In these models, the NLSP is typically either a bino or a stau, the former possibility giving rise to the $\gamma\gamma+\met$ signature.  As we will review below, strong bounds have been set on these models by the CDF and \dzero collaborations \citep{oldcdfgammagamma,Buescher:2005he,Abazov:2004jx,Abazov:2007is, CDFggmet}, and proposed searches in this channel have been put forth by the ATLAS  and CMS collaborations~\citep{ATLASGMSB,cmsgmsb}.  

In recent years, there have been a number of theoretical developments in SUSY-breaking and model building which have greatly expanded the phenomenological possibilities of gauge mediation.  For instance, in \citep{EOGM} the most general renormalizable, weakly-coupled extensions of ordinary gauge mediation were analyzed in detail. It was shown that these ``(extra)ordinary gauge mediation"  models could realize very different spectra than those of ordinary gauge mediation. In one especially interesting set of models, the NLSP was a Higgsino-like neutralino, instead of bino-like as in ordinary gauge mediation. Apart from arising in this simple class of models, Higgsino NLSPs are interesting in their own right, as they are generic features of gauge mediation models which solve the little hierarchy problem \citep{AG}. That is, the neutralino NLSP is mostly Higgsino-like in the limit of small $\mu$, and this is also the limit in which minimizes the fine-tuning associated with the Higgs sector in the MSSM.

In \citep{GGM1}, the question of how to formulate gauge mediation in a model-independent way was examined. This resulted in the framework of General Gauge Mediation (GGM), in which the entire space of possible gauge mediation models can be investigated without having to rely upon specific models. The basic idea here was to view the SM gauge group as a weakly-gauged global symmetry of the SUSY-breaking hidden sector, and to relate the MSSM soft masses in gauge mediation to correlation functions of the associated conserved currents. The  formulas obtained in this way are then exact in the hidden sector interactions but valid to leading order in the SM gauge couplings. Using this machinery, the complete set of independent parameters of gauge mediation can be derived, and in \citep{GGM1} they were shown to be: three arbitrary complex gaugino masses $M_{1,2,3}$; and three real parameters controlling the five sfermion mass-squareds $m_{Q,U,D,L,E}^2$. It was also argued that in the GGM framework, the $A$-terms are always small. The extension to the Higgs sector was considered in \citep{GGM1} and in much more detail in~\citep{Komargodski:2008ax}. Finally, in \citep{GGM2}, simple toy models were constructed which demonstrate that the entire parameter space of GGM can be covered. This opens the door to new avenues of gauge mediation phenomenology. In particular, it implies that any MSSM superpartner can be the NLSP in gauge mediation. For some recent work on the phenomenology of GGM, see \citep{Carpenter:2008he,Rajaraman:2009ga,Abel:2009ve}. Here let us briefly survey the possibilities. (For a similar survey that predates GGM and Tevatron Run II, see the last section of \citep{WorkingGroup}.)

If the NLSP is a colored sparticle, then the inclusive signature (assuming prompt decays) is dijets plus missing energy.  Therefore, the bounds on this scenario are likely to be similar to those found in searches for MSUGRA, approximately 300 GeV \citep{Aaltonen:2008rv}.  From the point of view of naturalness, it is somewhat unappealing to have the bottom of the MSSM sparticle spectrum be so high. Also, this is an extremely well-studied final state at the Tevatron, so there is not much new that can be added. 

Next we come to slepton NLSPs. An important point to keep in mind is that there is very little direct production of sleptons at the Tevatron or LHC. Instead, sleptons are more likely to be produced from cascade decays originating in pair production of colored sparticles or charginos and neutralinos. So  searches for sleptons will necessarily be more model dependent, involving more parameters of the spectrum. Furthermore, charged slepton NLSPs are already well-studied in the gauge mediation literature. Sneutrino NLSPs are an intriguing possibility which we will not discuss here; see however \citep{sneutrinonlsp}.

Finally, we arrive at neutralino and chargino NLSPs. Since these can be directly produced at the Tevatron, they are in some sense the minimal module for MSSM phenomenology at the Tevatron. However, charginos can only be NLSPs in narrow regions of parameter space, as we will review below, and as was analyzed recently in \citep{Kribs:2008hq}. Thus we are left with general neutralino NLSPs as the remaining possibility for interesting, new, generic phenomenology. 

Motivated by the recent progress and these considerations, we will examine in this paper the phenomenology of promptly-decaying general neutralino NLSPs at the Tevatron.   In general, the neutralino can be an arbitrary mixture of bino, wino and Higgsino gauge eigenstates. Therefore, having pairs of general neutralino NLSPs in every SUSY event will give rise to inclusive signatures involving any pairwise combination of $\gamma+\met$, $Z+\met$ or $h+\met$. While the $\gamma\gamma+\met$ channel has been extensively examined at the Tevatron, channels involving $Z$'s and $h$'s  have not been used to search for gauge mediation previously. (Prior to Tevatron Run II, the phenomenology of Higgsino-like NLSPs had been studied in detail \citep{WorkingGroup,MatchevThomas, BMTW,Dimopoulos:1996vz,Dimopoulos:1996fj}, but not much experimental work has followed up on this.) Additionally, if there is a small splitting between the neutralino NLSP and the lightest chargino, there can be decay chains which end with $W^\pm+\met$ which have been almost completely ignored in the gauge mediation literature.  In this paper we will analyze these cases and others in detail to give a more complete picture.

We choose to focus on the Tevatron and not the LHC in this paper, because the Tevatron is a well-understood, smoothly-running machine which already has a large recorded dataset of about 6.5 fb$^{-1}$ per experiment at the time of this writing (and more on the way). So it would be a shame not to push the existing Tevatron data as far as possible in the search for plausible new physics like gauge mediation. Also, using Tevatron data to constrain a large region of the parameter space of light neutralino NLSPs can help to focus future effort, both by model-builders and experimentalists. Of course, it is extremely important and interesting to ask how well the LHC can do, and how quickly it will surpass the capabilities of the Tevatron. These are issues that we will return to in a future publication \citep{gmsblhc}.  

In addition, the long-lived NLSP scenario is also clearly interesting. But since it involves qualitatively different physics and experimental techniques from the prompt case, we will examine it in a separate, forthcoming publication \citep{llgmsb}.

\subsection{Search strategies and analysis methodology}
\label{subsec:pgs}

In the following sections, we will re-analyze existing Tevatron searches in final states relevant to general neutralino NLSPs and derive their constraints on the  parameter space. We will also suggest, wherever possible, potential improvements to the existing searches that could enhance the sensitivity. Before moving on to the detailed discussion, let us first comment briefly on our general methodology.

We will be focusing mostly on counting experiments. This is because SUSY cross sections at the Tevatron are generally rather low, so we are usually dealing with small numbers of events, making it difficult to study the detailed signal shape. Furthermore, we will be focused on obtaining current or projected 95\% confidence-level exclusion limits, since these offer the most optimistic estimates for the ``reach" of the collider. However, wherever possible, we will also mention our estimates of the 5$\sigma$ discovery potential and the 3$\sigma$ ``evidence" threshold. We will use the maximum likelihood method to calculate the statistical significance of a given experimental search, following the discussion in \citep{Conway:2000ju}. For simplicity, we do not include systematic uncertainties in our estimates (in any event, we expect them to be small for signals involving leptons and photons). We also rely on tree-level estimates of signal rates. This is likely to lead to conservative estimates of the experimental sensitivity, because we are using accurate background rates, while the signal rates are likely to be enhanced significantly at NLO.

When calculating signal rates, we will be assuming that both the colored sparticles and the sleptons are decoupled, so that the signal cross section comes entirely from pair production of charginos and neutralinos.
Gluinos and squarks contribute significantly to the SUSY cross section at the Tevatron only when $m_{colored}\lesssim 450$ GeV. Since it is often the case that they are much heavier than this in models of gauge mediation, it is reasonable to neglect them. Similarly, the cross sections for direct slepton production at the Tevatron are small unless the sleptons are extremely light, so even if sleptons were present in the spectrum, they would have little effect on our conclusions. These assumptions simplify the analysis considerably (fewer parameters to consider), and lead to more conservative bounds.

All simulations discussed in the paper use Pythia \citep{Pythia} to generate signal and background samples, except for a few backgrounds with high-multiplicity final states (e.g. $Zb\bar{b}$ and $b\bar{b}b\bar{b}$ in section \ref{sec:bjets}) for which we have used MadGraph \citep{MadGraph} to obtain the correct tree-level matrix elements (and, subsequently, Pythia to shower and hadronize the parton-level events). These samples were run through the PGS detector simulation \citep{PGS}, which includes models of tracking efficiency, calorimeter response, and $b$-tagging. We have made small modifications of PGS to better account for the detector geometry and cracks at CDF and \dz. Wherever possible, we have estimated backgrounds using PGS and compared them to published measurements and CDF or \dzero background estimates that were computed with the full detector simulation and/or derived from data control samples. In channels involving real identified objects (leptons, photons) and real missing $E_T$, we find that PGS is typically accurate at the 20\% level, and is often much better. This gives us confidence that our signal simulations (which also involve real objects and real $\met$, of course) also achieve this level of accuracy. We find that PGS is less reliable when estimating backgrounds involving fake objects or fake missing $E_T$; for these backgrounds, we will rely on published experimental results as much as possible. We further note that since the signal cross sections times branching ratios are generally rapidly falling functions of the sparticle masses, even a large error in the simulation of detector response can lead to a relatively small error in the estimated reach in mass.

\subsection{Relation to previous work}

In many respects, the work presented here revisits topics previously covered over a decade ago by many authors \citep{WorkingGroup,MatchevThomas, BMTW,Dimopoulos:1996vz,Dimopoulos:1996fj} in preparation for Tevatron Run II. These important and comprehensive papers also studied the collider phenomenology of general neutralino NLSPs (along with many other cases), and they also used Monte Carlo simulations to examine the discovery potential in all the various final states.  Since there is significant overlap with these older papers, it is worth highlighting briefly the ways in which our work here differs or goes beyond the previous literature.

First and foremost, our reliance on detailed background estimates from both data and Monte Carlo, and on the established capabilities of the Tevatron detectors in Run II, distinguishes our analysis from earlier discussions of Higgsino NLSPs. Since these predated Run II, they were necessarily of a more rough and speculative nature. 

Second, we attempt to phrase things in a completely model independent fashion, in terms of the general neutralino parameter space (to be described later). This is in contrast with the old works which focused on a few ``model lines"  -- one-dimensional slices through parameter space. 

Third, we also consider final states that were overlooked or discussed only briefly in the earlier literature; in particular, our analysis of wino co-NLSPs is an important addition.

Finally, one of the main points of our paper is that -- with the exception of $\gamma\gamma+\met$ -- essentially none of the searches proposed in the old literature have been explicitly followed up on by experimentalists. At the same time, however, there have been searches in many of the important final states, either in a ``signature-based" fashion, or motivated by  scenarios other than gauge mediation. So in the present work, we hope to emphasize that experimentalists should be able to easily adapt existing searches to constrain an important portion of gauge mediation parameter space. Moreover, by optimizing these searches for general neutralino NLSPs, they may even have the possibility of discovery or 3$\sigma$ evidence.

\subsection{Outline}

The outline of the paper is as follows.  In Section~\ref{sec:parspace} we review the dominant decay modes of general neutralino NLSPs. We also examine in detail the special simplifying limits in which the neutralino becomes a pure gauge eigenstate -- bino, wino or Higgsino NLSPs. In Section~\ref{sec:searches} we analyze the phenomenology of neutralino NLSPs at the Tevatron -- production cross sections, the most promising final states (in terms of $S/\sqrt{B}$), and the methodology we employ to analyze these channels. In section 4 we will discuss in detail the $\gamma\gamma+\met$ search and its applications to general neutralino NLSPs (including bino NLSPs, of course).  In sections 5-7, we will discuss searches relevant for wino, $Z$-rich Higgsino, and $h$-rich Higgsino NLSPs, respectively.   In Section 8 we summarize our results and conclude with some directions for future research, including comments about general neutralino NLSPs at the LHC.

\section{Review of Neutralino NLSPs}
\label{sec:parspace}
\setcounter{equation}{0}

\subsection{General neutralino NLSPs}

In this section we will review the different types of neutralino NLSPs and their decay modes. As discussed in the introduction, the general neutralino NLSP can be any linear combination of bino, wino, and Higgsino gauge eigenstates:
\beq
\tilde\chi_1^0 = \sum_{i=1}^4 N_{1i}\tilde\psi_i^0
\eeq
where $\tilde\psi_i^0=(\tilde B,\tilde W,\tilde H_d^0,\tilde H_u^0)$. The mass eigenvectors $N_{1i}$ depend on four MSSM parameters: $M_1$, the soft mass for the bino; $M_2$, the soft mass for the wino; $\mu$, the supersymmetric Higgs mass term; and $\tan \beta$, the ratio of the up-type to down-type Higgs VEVs. We assume the mass parameters are all real to avoid the SUSY CP problem; however, they can have either sign. (For more detailed discussion of the neutralino spectrum and parameter space, see e.g.\ Martin's excellent review of the MSSM \cite{Martin:1997ns}.)  

The general neutralino NLSP has three possible decay modes: gravitino + ($\gamma/Z/H$). As the parameters ($M_1,\,M_2,\,\mu,\,\tan\beta$) are changed, different decay modes will dominate.  The general formulas for the decay widths are \citep{Ambrosanio:1996jn,Dimopoulos:1996yq}:
\begin{eqnarray}
\label{eqn:decaywidths}
\Gamma(\tilde{\chi}^0_1 \rightarrow \tilde{G} + \gamma) & = &  \left|N_{11} c_W
 + N_{12} s_W\right|^2  {\cal A} \nonumber \\
\Gamma(\tilde{\chi}^0_1 \rightarrow \tilde{G} + Z) & = & \left(\left|N_{12} c_W
 - N_{11} s_W\right|^2 +{1\over2} \left|N_{13} c_\beta - N_{14} s_\beta \right|^2 \right)
 \left(1 - \frac{m_Z^2}{m^2_{\tilde{\chi}^0_1}}\right)^4 {\cal A} \nonumber \\
\Gamma(\tilde{\chi}^0_1 \rightarrow \tilde{G} + h) & = & {1\over2}\left|N_{13} c_\beta + N_{14} s_\beta\right|^2 \left(1 - \frac{m_h^2}{m^2_{\tilde{\chi}^0_1}}\right)^4 {\cal A}
\end{eqnarray}
These formulas are valid when the Higgs and $Z$ decays are on-shell. (When necessary, we will also employ the full expressions including finite-width effects, which can be found in \citep{BMPZii}.) In the third equation, we have also assumed the SM Higgs decoupling limit. ${\cal A}$ is a dimensionful parameter that sets the overall scale of the NLSP lifetime:
\begin{equation}
\label{eqn:decaylength}
{\cal A} = \frac{m^5_{\tilde{\chi}^0_1}}{16\pi F_0^2} \approx \left(\frac{m_{\tilde{\chi}^0_1}}{100~{\rm GeV}}\right)^5 \left(\frac{100~\rm{TeV}}{\sqrt{F_0}}\right)^4 \frac{1}{0.1~{\rm mm}}.
\end{equation}
Here $\sqrt{F_0}$ is the fundamental scale of SUSY breaking; it is related to the gravitino mass via $m_{3/2} = F_0/(\sqrt{3}M_{Pl})$. The range of possible $\sqrt{F_0}$ is roughly
\beq
\label{eq:Frange}
10\,\,{\rm TeV} \lesssim \sqrt{F_0} \lesssim 10^6\,\,{\rm TeV}
\eeq
where the lower bound comes from the requirement of  a viable superpartner spectrum, and the upper bound comes from requiring gauge mediation to dominate over gravity mediation. This spans the range from prompt decays to lifetimes significantly longer than the size of the detector. Indeed, for much of the window (\ref{eq:Frange}), i.e.\ if $\sqrt{F_0}$ is somewhat large compared to 100 TeV, neutralino decays are macroscopically displaced.  As discussed in section \ref{sec:intro}, in this paper we will only focus on the case where the decay of the NLSP is prompt.

Since the complete neutralino parameter space is four dimensional, it can be difficult to discuss the collider phenomenology in full generality. Instead, in this paper, we will mostly be studying various simplifying limits where the neutralino becomes a pure gauge eigenstate, i.e.\ where it becomes predominantly bino, wino, or Higgsino-like. For much of the parameter space, this is a pretty good approximation. Moreover, each limit has distinctive phenomenology. By considering each in turn, we capture the essential physics of general neutralino NLSPs.

\subsection{Bino NLSP}

The neutralino NLSP becomes bino-like in the limit $|N_{11}|\gg |N_{12}|,|N_{13}|,|N_{14}|$. In terms of the MSSM parameters, this corresponds to $|M_1|\ll |\mu|$ and $|M_1|<|M_2|$. Substituting $N_{11}\to 1$ into (\ref{eqn:decaywidths}), we find 
\begin{eqnarray}
\label{eqn:binodecay}
\Gamma(\tilde\chi_1^0\to \tilde G+\gamma)&=&c_W^2{\mathcal A}\nonumber\\
 \Gamma(\tilde\chi_1^0\to \tilde G+Z)&=&s_W^2\left(1 - \frac{m_Z^2}{m^2_{\tilde{\chi}^0_1}}\right)^4 {\cal A}\\
  \Gamma(\tilde\chi_1^0\to \tilde G+h)&=&0
\end{eqnarray}
with $m_{\tilde\chi_1^0}\approx |M_1|$. So we see that the bino-like NLSP decays to photons at least $c_W^2\approx 76\%$ of the time, and to $Z$ bosons the rest of the time. The branching fraction to photons can be further enhanced for lighter NLSPs, due to the $m_Z$ phase space suppression factor.

\subsection{Wino NLSP}

The neutralino NLSP becomes wino-like in the limit $|N_{12}|\gg |N_{11}|,|N_{13}|,|N_{14}|$.  This corresponds to $|M_2|\ll \mu$ and $|M_2|<|M_1|$. In this regime, the decay widths are
\begin{eqnarray}
\label{eqn:winodecay}
\Gamma(\tilde\chi_1^0\to \tilde G+\gamma)&=&s_W^2{\mathcal A}\nonumber\\
 \Gamma(\tilde\chi_1^0\to \tilde G+Z)&=&c_W^2\left(1 - \frac{m_Z^2}{m^2_{\tilde{\chi}^0_1}}\right)^4 {\cal A}\\
  \Gamma(\tilde\chi_1^0\to \tilde G+h)&=& 0
\end{eqnarray}
with $m_{\tilde\chi_1^0}\approx |M_2|$. Here the NLSP decays to photons at least 23\% of the time, and to $Z$s the rest of the time. As with bino NLSPs, the branching fraction to photons can be enhanced if the NLSP is light, due to the $Z$-boson phase space factor. Indeed, for NLSP masses less than 150 GeV, the branching fraction to photons is greater than 50\%. 

One special feature in the wino NLSP limit is that the lightest chargino -- whose mass is also approximately $|M_2|$ here -- is generically a co-NLSP. The decay rate of wino-like charginos to $W^+$ plus gravitino is given by \citep{Ambrosanio:1996jn}
\beq
\label{eqn:chWG}
\Gamma(\tilde \chi^+_1 \to W^+ + \tilde{G}) &=& {1\over2}\left(1-{m_W^2\over m_{\tilde\chi^+_1}^2}\right)^4 \CA 
\eeq
If the chargino-neutralino mass splitting is sufficiently small, 
\beq
\label{eqn:conlspcrit}
\Delta m \equiv |m_{\tilde\chi^+_1}-m_{\tilde\chi^0_1}| \lesssim 1\,\,{\rm GeV} 
\eeq
then the decay length of $\tilde\chi^+_1 \to \tilde\chi^0_1 + W^*$ is longer than $\sim$ 0.1 mm and the decay to gravitino (assumed here to be prompt) dominates \citep{CDG1,CDG2,CDM}. When this is the case, we will refer to the wino-like chargino as a co-NLSP.\footnote{We note that regardless of whether it is a co-NLSP, the lightest chargino will {\em always} decay promptly as long as the neutralino does as well.  This should be contrasted with other scenarios with degenerate winos in the MSSM, for instance as in AMSB~\citep{AMSBphen}.}

As is well known \cite{Martin:1993ft}, the splitting for wino charged and neutral mass eigenstates is abnormally small, due to a cancellation at leading order in $m_Z/\mu$.  The tree-level splitting goes like $\Delta m \sim m_Z^4/\mu^3$, so as long as $\mu$ is moderately large ($\mu\gtrsim$ a few hundred GeV), the splittings will be sub-GeV. For this reason, when discussing the collider signatures of winos below, we will always assume that they are co-NLSPs.

\subsection{Higgsino NLSP}
\label{subsec:HiggsinoNLSP}

Higgsino NLSPs have been studied previously in \citep{MatchevThomas, BMTW,BMTWLHC,Dimopoulos:1996vz,Dimopoulos:1996fj}.  Here we will review their basic features, with a greater emphasis towards the model-independent parameter space. The neutralino NLSP becomes Higgsino-like in the limit $|N_{13}|,|N_{14}|\gg |N_{11}|,|N_{12}|$, which corresponds to $|\mu|\ll |M_1|,\, |M_2|$. In fact, this limit includes two distinct possibilities: 
\beq
N_{13}=-\eta N_{14}={1\over\sqrt{2}}
\eeq
with $\eta=\pm1$.\footnote{In terms of the underlying MSSM parameters, one finds $\eta \equiv {\rm sign}(\mu)\times{\rm sign}\left({M_1\over M_2}+\tan^2\theta_W\right)$. While this depends on the relative sign between $M_1$ and $M_2$, in this paper we will only analyze the case that $M_1,M_2>0$.  This is sufficient to capture the phenomenology of all neutralino NLSPs, and in this case, $\eta$ reduces to the more familiar variable ${\rm sign}(\mu)$ used in many phenomenological analyses.}
The decay widths are given by
\begin{eqnarray}
\label{eqn:brfhi}
\Gamma(\tilde\chi_1^0\to \tilde G+\gamma)&=& {1\over2}(s_\beta+\eta c_\beta)^2\left({c_Ws_W(M_1-M_2)m_Z\over M_1 M_2}\right)^2\CA
\nonumber\\
 \Gamma(\tilde\chi_1^0\to \tilde G+Z)&=&{1\over4}(s_\beta+\eta c_\beta)^2\left(1 - \frac{m_Z^2}{m^2_{\tilde{\chi}^0_1}}\right)^4 {\cal A}\\
  \Gamma(\tilde\chi_1^0\to \tilde G+h)&=&{1\over4}(s_\beta-\eta c_\beta)^2\left(1 - \frac{m_h^2}{m^2_{\tilde{\chi}^0_1}}\right)^4 {\cal A}\nonumber
\end{eqnarray}
In the first equation, we have kept the leading-order term in the large $M_{1,2}$ limit, since this can affect the branching fractions at small NLSP mass where the $Z$ and $h$ decays are strongly phase-space suppressed.

\begin{figure}[t!]
\begin{center}
\includegraphics[width=0.9\textwidth]{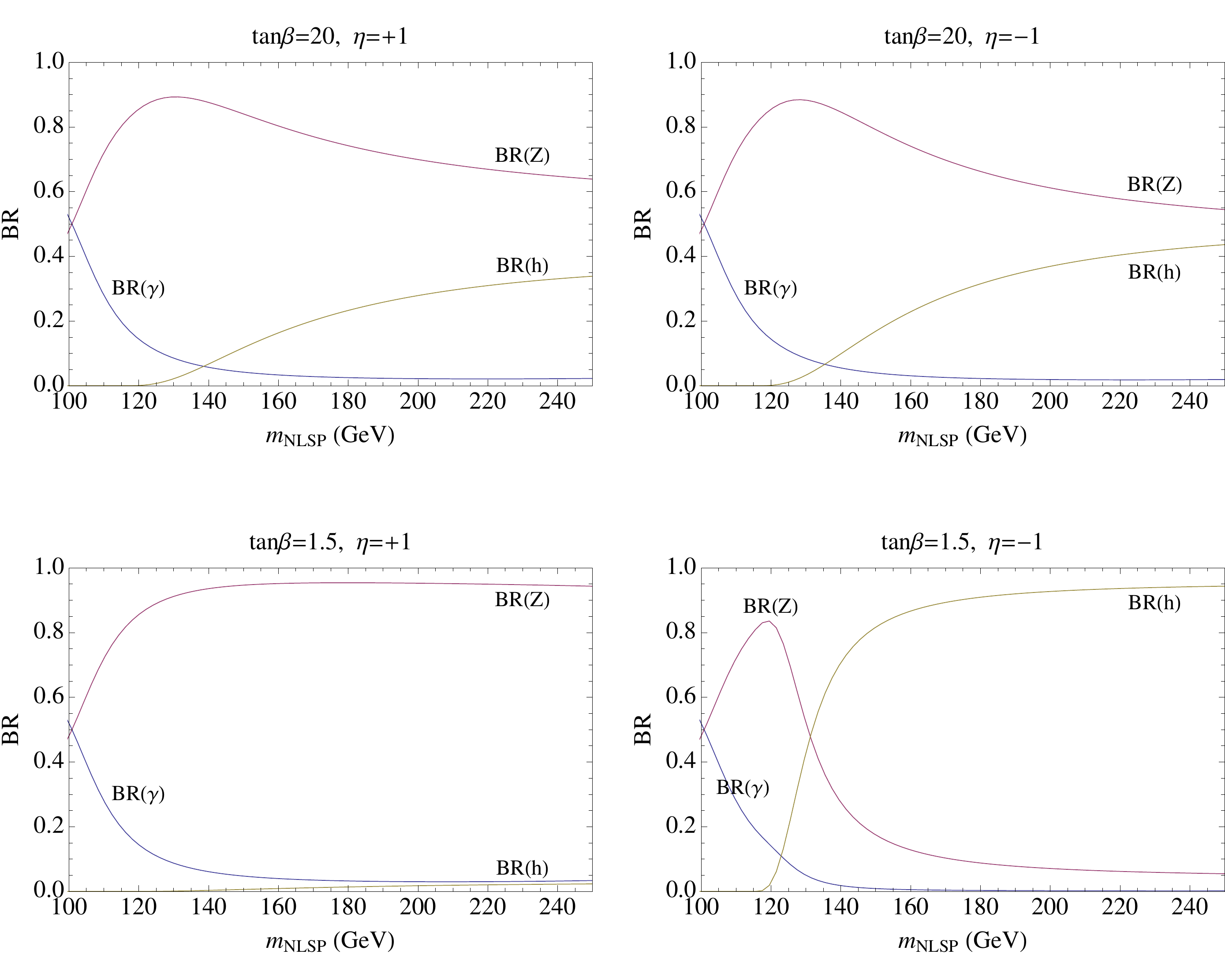}
\end{center}
\caption{Branching ratios of the Higgsino NLSP to photons, $Z$'s and Higgses, as a function of $m_{NLSP}$, for $\eta=\pm1$ and $\tan\beta=1.5,\,20$. In all the plots, $M_1=500$ GeV, $M_2=1000$ GeV, and $m_{h^0}=115$ GeV. }
\label{fig:HiggsinoBR}
\end{figure}

In figure \ref{fig:HiggsinoBR} we plot the branching fractions for different values of $\tan\beta$ and $\eta$. (For the sake of argument, we have taken $m_h=115$ GeV in all these plots, and also for the remainder of the paper. Similar plots and discussion of the branching fractions can also be found in \cite{MatchevThomas}.)   We see that for low mass NLSPs, there can be a significant branching fraction to photons, due to the phase space suppression factors (or simply being kinematically disallowed) in the other channels. Here we should emphasize that $\Br(\gamma)$ at low mass depends sensitively on $M_1$ and $M_2$, through (\ref{eqn:brfhi}). For intermediate mass NLSPs, the decay is dominantly to $Z$s. Finally, for high mass NLSPs, the decay is still dominantly to $Z$s, except for $\eta=-1$ and small $\tan\beta$, where it is dominantly to Higgses. (For extremely high masses, one can also have $\Br(Z)\approx \Br(h)$, but unfortunately this regime is inaccessible at the Tevatron.) Given the different possible behaviors of the branching fractions, we will find it useful in this paper to distinguish between photon-rich, $Z$-rich and Higgs-rich Higgsino NLSPs.

Similar to the wino NLSP, the lightest neutralinos and chargino $\tilde\chi_{1}^0$, $\tilde\chi_2^0$ and $\tilde\chi_1^\pm$ all have masses $\approx |\mu|$ in the Higgsino NLSP regime. However, the difference here is that in practice one typically needs unrealistically large values of $M_1$, $M_2$ in order for them to be co-NLSPs.  This is because (in contrast to the wino case) the splittings come at leading order in the small $m_Z$ expansion, i.e.\ $\Delta m \sim m_Z^2/M_{1,2}$. So in order for $\Delta m\lesssim 1$ GeV, one needs $M_{1,2}\gtrsim 10$ TeV.  For more realistic values of $M_1$ and $M_2$, the splittings are large enough that the heavier Higgsino-like states can decay promptly to the NLSP via off-shell $W$'s and $Z$'s going to jets or leptons. So we will not consider the possibility of Higgsino co-NLSPs in this paper.

Another possibility that we will briefly mention is that of the chargino NLSP. This occurs in a special corner of the Higgsino limit, where $\mu<0$ and $-\left({1+s_{2\beta} \over 1-s_{2\beta}}\right)t_W^2<  {M_1\over M_2} < -\left({1-s_{2\beta} \over 1+s_{2\beta}}\right)t_W^2$. The phenomenology of chargino NLSPs has recently been discussed in \citep{Kribs:2008hq}.  Since we are focusing on neutralino NLSPs in this paper,  we will not discuss this scenario further. 

\section{Overview of Tevatron Phenomenology}
\label{sec:searches}
\setcounter{equation}{0} 

\begin{figure}[!t]
\begin{center}
\includegraphics[width=\textwidth]{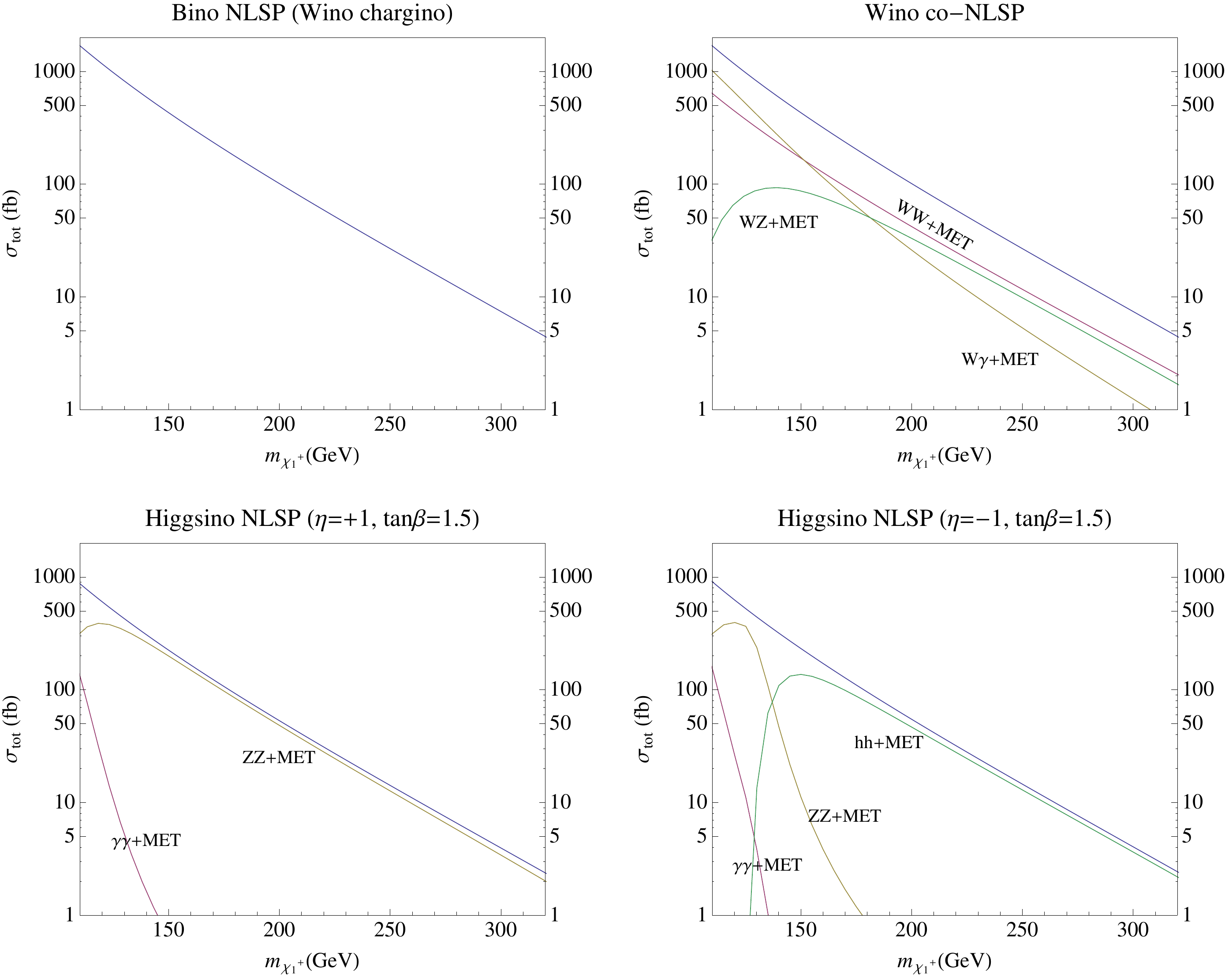}
\end{center}
\caption{Cross sections for bino, wino and Higgsino NLSPs, as a function of the chargino mass. In the case of binos, the cross section shown is for production of a wino state above the bino, because direct bino production is negligible. Shown also are the rates for various final states, obtained by multiplying the cross sections by the relevant branching ratios from section \ref{sec:parspace}. In order to not clutter the plots, we have not included the rates for mixed final states, as these are generally small and will be irrelevant to our analysis.} 
\label{fig:genxsecs}
\end{figure}

In this section, we will give a broad overview of the phenomenology of neutralino NLSPs at the Tevatron, starting with their production cross sections. NLSPs  are produced in every SUSY event, either directly or at the bottom of any decay chain. Thus the rate for NLSP  production is just given by the total SUSY cross section times the relevant branching ratios. As discussed in the introduction, we are assuming that the SUSY cross section at the Tevatron is given entirely by the chargino/neutralino sector.  Shown in fig.\  \ref{fig:genxsecs} are the total cross sections for bino, wino and Higgsino NLSPs, as well as the rates in various final states determined by the decay modes of the NLSP. Note that in the case of binos, the dominant cross section is not for direct production of binos themselves (this proceeds only through mixings), but for heavier Higgsino or wino states which then decay down to bino. The plotted cross sections and other estimates discussed in this section do not fold in geometric acceptances, except where otherwise noted. (Such acceptances will, of course, be included in the more detailed simulations in the remainder of the paper.) 

We clearly see from fig.\ \ref{fig:genxsecs} that regardless of the final state, the Tevatron will not be able to probe chargino masses above $300$ GeV. (Of course, this depends on the assumption of decoupled colored sparticles.)  Moreover, to probe this entire range one needs to have high signal acceptance and/or no branching fraction suppressions, as well as to be able to suppress the SM background entirely. As we will see below, only the bino-like NLSP has a chance of satisfying all these conditions, through the $\gamma\gamma+\met$ final state.

Now let us look at the final states for the different types of neutralino NLSPs and discuss which are most relevant. In the case of bino NLSPs, it is clear that an inclusive $\gamma\gamma + \met$ signature is the best search channel. The branching fraction to photons is large, photons are clean experimental objects, and simple cuts in this channel can eliminate essentially all SM backgrounds. This channel is useful even when the bino content is highly mixed with Higgsino, as we will show in section \ref{sec:phosearches}.

For wino and Higgsino NLSPs, there are many more final states to consider, involving photons, $W$'s, $Z$'s, and Higgses in various combinations. We will focus mostly on final states involving energetic photons and/or leptons, as these are clean experimental objects with small, well-characterized backgrounds. The one exception is for Higgs-rich Higgsino NLSPs, where we will also consider all-hadronic final states consisting of some number of  $b$-jets.

 For the case of wino co-NLSPs, production of $\tilde{\chi}^\pm_1 \tilde{\chi}^0_1$ leads to final states with $W^\pm\gamma + \met$ and $W^\pm Z +\met$, while  $\tilde{\chi}^+_1 \tilde{\chi}^-_1$ gives $W^+W^- + \met$.  (Direct production of $\tilde\chi_1^0\tilde\chi_1^0$ is negligible, since this proceeds only through mixing with the heavy Higgsinos.) The following simple $S/\sqrt{B}$ ``rules of thumb" indicate that only the first is a viable search channel for wino co-NLSPs: 
 
\begin{itemize}

\item{$W(\ell\nu)\gamma+\met$.} Here the irreducible SM background is $W(\ell\nu)+\gamma$. The cross section for this is a rapidly falling function of photon $p_T$; for $p_T(\gamma)>20$ GeV, one finds $\sigma\sim 1500$ fb at the Tevatron. As can be seen from the upper right plot in fig.\ \ref{fig:genxsecs}, after multiplying by $\Br(W\to\ell\nu)\approx 20\%$, the signal rate in this final state ranges from 200-10 fb for $m_{NLSP}$ between 100-200 GeV. So $S/\sqrt{B}\gtrsim 1$ for a sizeable range of NLSP mass, and  this is a promising channel in which to search for wino co-NLSPs.

\item{$W(\ell\nu)Z(\ell\ell)+\met$.} Here the irreducible background is simply SM $WZ$ production, whose rate in the trilepton plus $\met$ final state is $\sim 30$ fb at the Tevatron. Meanwhile,  fig.\ \ref{fig:genxsecs} shows that the signal rate in the $W(\ell\nu)Z(\ell\ell)+\met$ final state peaks at $\sim 1$ fb around $m_{NLSP}\sim 140$ GeV. So $S/\sqrt{B}$ is never $\CO(1)$ here, and this is not a promising channel. 

\item{$W({\rm jets})Z(\ell\ell)+\met$.} This final state has a signal rate that peaks at $\sim 4$ fb also at $m_{NLSP}\sim 140$ GeV. There are no irreducible SM backgrounds here; the dominant backgrounds turn out to be $Z+{\rm jets}$ and $t\bar t$. However if we factor in a signal acceptance that is at most 30\% (from a simulation of the $\eta$ distribution and an estimate of the geometric acceptance), there simply will not be enough events in this final state to offer any sensitivity.

\item{$W(\ell\nu)W(\ell\nu)+\met$.} The irreducible background for this final state is SM $WW$ production. The rate for this in the $\ell\ell+\met$ final state is $\sim 300$ fb. According to  fig.\ \ref{fig:genxsecs}, the signal goes from 25-2 fb as $m_{NLSP}$ goes from 100-200 GeV. Factoring in a 30\% geometric acceptance, we can barely get $S/\sqrt{B}\sim 2$ in 10 fb$^{-1}$. So this is also not a promising channel to search for wino NLSPs.

\end{itemize}

In the case of Higgsinos, we have the three different cases discussed in section \ref{subsec:HiggsinoNLSP}: photon-rich, $Z$-rich and Higgs-rich Higgsino NLSPs. The relevant final states were considered in \citep{MatchevThomas, BMTW,Dimopoulos:1996vz,Dimopoulos:1996fj}, and here we will review them all. 

For the photon-rich case (low-mass Higgsino NLSPs with moderate $M_1$ and $M_2$), the best channel is again obviously $\gamma\gamma+\met$. 

For the $Z$-rich case, the leptonic decay of the $Z$ boson provides us with clean final states with tractable backgrounds, an advantage that outweighs its small branching fraction. The possible final states are $Z(\ell^+\ell^-)Z({\rm jets})+\met$ and $Z(\ell^+\ell^-)Z(\ell'^+\ell'^-)+\met$. Rough $S/\sqrt{B}$ estimates indicate that both channels could be useful:

\begin{itemize}

\item{$Z(\ell^+\ell^-)Z({\rm jets})+\met$.} Here there are essentially no irreducible SM backgrounds, only reducible ones such as $Z+{\rm jets}$, $t\bar t$ and dibosons. The signal rate is sufficiently high, ranging from 30-5 fb for $m_{NLSP}$ between 100-200 GeV, according to the lower left plot in fig.\ \ref{fig:genxsecs}. So this channel is a promising one.

\item{$Z(\ell^+\ell^-)Z(\ell'^+\ell'^-)+\met$}. This has no SM irreducible background, but the signal rate is small due to the $\Br(Z\to\ell^+\ell^-)^2$ suppression. It ranges from 1.5-0.2 fb for $m_{NLSP}$ between 100-200 GeV, according to fig.\ \ref{fig:genxsecs}. So the channel is clean, but whether one can discover anything here depends on how high the signal acceptance can be made.

\end{itemize}

For the Higgs-rich Higgsino NLSP in the $hh+\met$ final state, the lower right plot in fig.\ \ref{fig:genxsecs} indicates that the total cross section peaks at $\sim 150$ fb at $m_{NLSP}\sim 140$ GeV, and it decreases to $\sim 50$ fb at $m_{NLSP}\sim 200$ GeV. By comparison, the SM $(W/Z)+h$ rate is only $\sim 70$ fb at the Tevatron for $m_h\approx 115$ GeV, after factoring in the leptonic branching fractions of the $W$ and $Z$.  Since this is the preferred search mode for a light Higgs at the Tevatron, we see that the Higgs-rich Higgsino NLSP can have comparable rates to the SM Higgs at the Tevatron. This offers an extremely interesting possibility to discover not only BSM physics, but also the light Higgs at the Tevatron! 

Here the backgrounds are much more difficult to estimate, so we will not attempt any $S/\sqrt{B}$ comparisons. Instead we will only enumerate the possible final states and estimate the signal rates. The final states are dictated by the decay modes of the Higgs. Assuming a light Higgs ($m_h\sim 115$ GeV), the largest are \cite{Carena:2000yx}: $\Br(h\to b\bar b)\sim 70\%$,  $\Br(h\to WW^*)\sim 10\%$, $\Br(h\to\tau^+\tau^-)\sim 7\%$ (we are ignoring $h\to gg$). From this, it is clear that to get an appreciable signal rate, we can only look at inclusive final states involving $b$-jets. If we require at least two tagged $b$-jets in the final state, then using a $\sim 50\%$ $b$-tagging efficiency, the signal rate becomes 10-30 fb. The $t\bar{t}$ cross section is 7.5 pb, and cross sections for $Wb\bar{b}$, $Zb\bar{b}$, and diboson production are also of this order. Multi-jet backgrounds with fake $\met$ or $\met$ arising from semileptonic $b$ or $c$ decays are also important. Clearly, shape information and not just counts will be important for reducing backgrounds. Requiring more $b$-jets reduces the signal rate accordingly, but presumably reduces the SM background by an even larger factor. So perhaps this is a promising way to search for Higgs-rich Higgsino NLSPs.

Finally, let us briefly mention the possibility of mixed final states: $\gamma+Z+\met$, $\gamma+h+\met$, or $Z+h+\met$.  Because $\Br(Z\to \ell^+ \ell^-) \approx 6\%$ and photon channels are very clean, the $\gamma+Z$ channel is never preferable to $\gamma+\gamma+\met$ in the region where the photon branching fraction is significant. Furthermore, it is clear that $\gamma+h$ can only be useful in a very small region of parameter space where $\eta = -1$ and $\tan \beta \approx 1$ (see the plots in Figure \ref{fig:HiggsinoBR}). We will comment briefly on this scenario in Section \ref{sec:bjets}; it suffers from large QCD backgrounds. Finally, an inclusive search strategy for $Z(\to \ell^+ \ell^-) + \met + X$ will do just as well for the mixed $Z+h$ case as for $Z(\ell^+\ell^-)Z({\rm jets})+\met$ (one could inspect additional jets for $b$-tags and mass peaks, though in practice the reach is too limited for this to be of much interest).

To summarize, we have seen that the best channels to search for the various types of neutralino NLSPs are: $\gamma\gamma+\met$ for bino NLSPs and photon-rich Higgsino NLSPs; $W(\ell\nu)+\gamma+\met$ for wino co-NLSPs; $Z(\ell\ell)+\met+X$ and $Z(\ell\ell)+Z(\ell'\ell')+\met$ for $Z$-rich Higgsino NLSPs; and multi-$b$-jets plus $\met$ for $h$-rich Higgsino NLSPs. In the following sections, we will apply existing Tevatron analyses in these final states to constrain general neutralino NLSPs and describe how these analyses can be improved to optimize the sensitivity. These sections are organized around the different final states, so the reader who is interested in only specific final states can skip to the relevant (sub)sections.

\section{Searches for $\gamma\gamma+\met$ and their Applications}
\label{sec:phosearches}
\setcounter{equation}{0} 

The classic search for gauge mediation is the diphoton plus missing energy signature. It is a clean channel, with SM backgrounds arising primarily from QCD processes ($\gamma\gamma$, $\gamma j$, and $jj$) with fake $\met$, and from electroweak processes ($W/Z+\gamma\gamma$; and $W/Z+\gamma+j$, $W/Z+j$) with real $\met$. This channel is primarily useful for constraining bino-like NLSPs, in which the branching fraction to photons is guaranteed to be large.   The signal rate is essentially that shown in fig.\ \ref{fig:genxsecs}, and the SM irreducible backgrounds are virtually nonexistent. So this is an extremely promising channel, where we are only limited by the signal acceptance.

There are numerous Tevatron searches in this channel \citep{oldcdfgammagamma,Buescher:2005he,Abazov:2004jx,Abazov:2007is, CDFggmet}, the most recent of which comes from CDF with 2.6 fb$^{-1}$ integrated luminosity \citep{CDFggmet}. This search required two central 13 GeV photons which are not back-to-back, $H_T$ of 200 GeV, and $\met$ of 3-sigma significance. Zero events were found with an expected background of 1.2 events. 
This implies the following 95\% confidence limit on the cross section times branching ratio for general neutralino NLSPs:
\beq
\label{eqn:ggmetbound}
\sigma_{tot}\times \Br(\gamma)^2 \times \varepsilon \lesssim 1.2\,\,{\rm fb}
\eeq
where $\varepsilon$ is the signal acceptance of the CDF analysis. It is also trivial to compute the projected 95\% confidence limit that will be set by the CDF $\gamma\gamma+\met$ analysis with 10 fb$^{-1}$ of integrated luminosity:
\beq
\label{eqn:ggmetprojbound}
\sigma_{tot}\times \Br(\gamma)^2 \times \varepsilon \lesssim 0.55\,\,{\rm fb}\qquad (10\,\,{\rm fb}^{-1})
\eeq
One can check that  with 10 fb$^{-1}$ of data, the $\gamma\gamma+\met$ channel does not offer any opportunity for 5$\sigma$ discovery.

\begin{figure}[!t]
\begin{center}
\includegraphics[width=\textwidth]{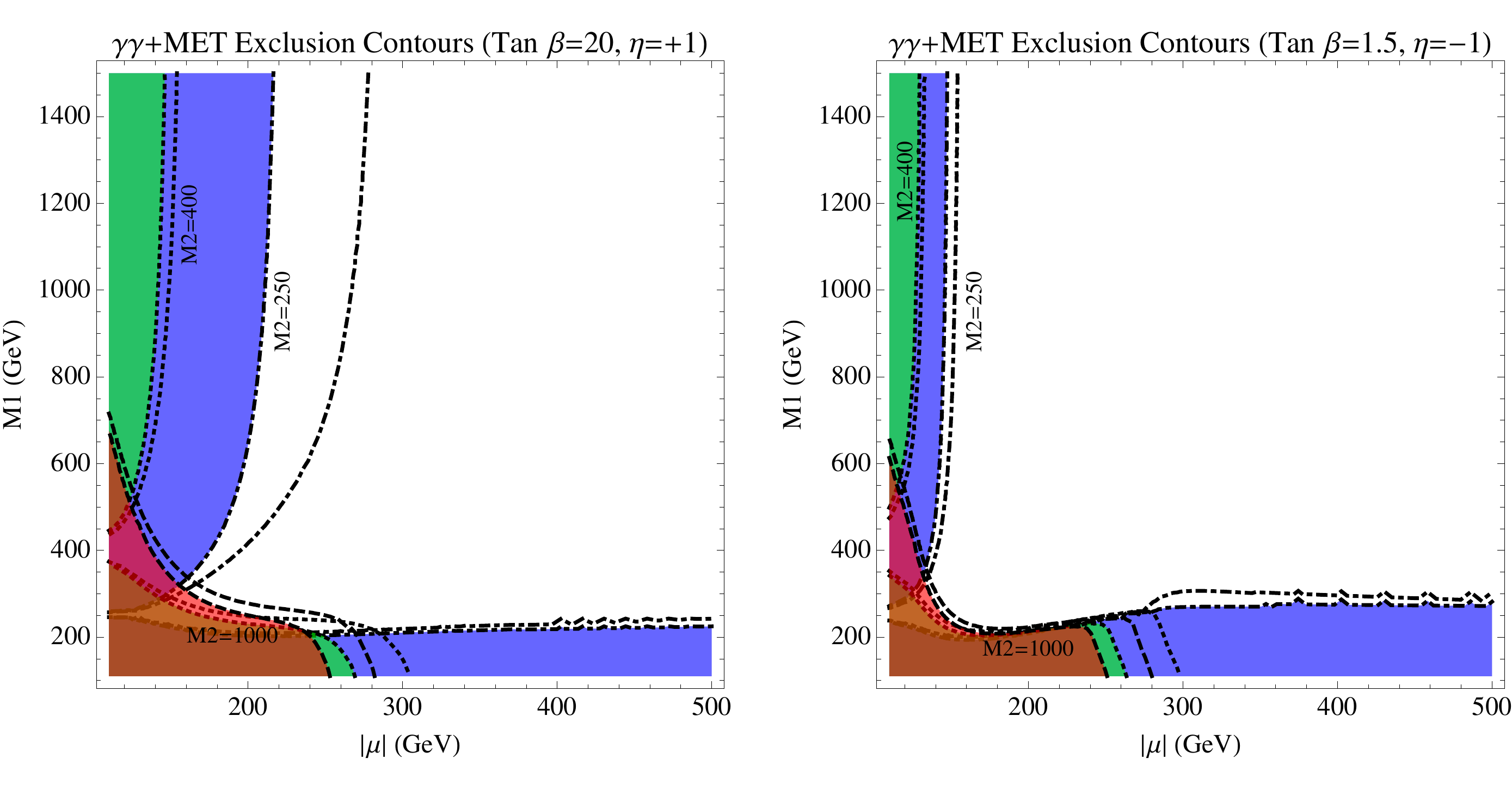}
\end{center}
\caption{Various examples of 95\% exclusion contours (shaded regions) in the $(\mu, M_1)$ plane, derived from (\ref{eqn:ggmetbound}), for the most recent 2.6 fb$^{-1}$ CDF search for anomalous $\gamma\gamma+\met$ production. Also shown are the projected exclusion contours with 10 fb$^{-1}$ of data (dashed lines). On the left: exclusion contours for $\tan \beta = 20$, $\eta=+1$ and various values of $M_2$. On the right: the same, but for $\tan\beta=1.5$, $\eta=-1$.} 
\label{fig:exclggmetgen}
\end{figure}

In fig.\ \ref{fig:exclggmetgen}, examples of exclusion contours are shown in the ($\mu$, $M_1$) plane, for various choices of $M_2$, $\tan\beta$ and $\eta$. These plots, unlike the remainder of the plots in this paper, are based solely on analytic calculations of the cross sections, rather than a Pythia and PGS simulation, because here we are able to rely on the experiment's quoted value of $\varepsilon$.\footnote{To make the exclusion plots, we have assumed a constant overall acceptance of $\varepsilon=7.8\%$, as quoted in the CDF analysis for the MGM benchmark model SPS8 \citep{SPS}. In reality, the acceptance varies somewhat as we move about in the parameter space, due mostly to the dependence of the photon isolation requirements on the splittings between different -ino states. In our PGS simulations, we have found variations by as much as a factor of 2 from the efficiency quoted in the CDF study. However, because $\sigma \times Br$ is varying rapidly across the parameter space, we find that even if our estimate of the efficiencies is off by a factor of 2, our reach estimates vary by less than 5\%.} 
These plots illustrate the power of $\gamma\gamma+\met$ in constraining the general neutralino NLSP parameter space, and not just the specific simplifying limits described in the previous sections. Taking $M_2=1000$ GeV illustrates how mixed bino-Higgsinos can be excluded with $\gamma\gamma+\met$. Similarly, the curves with $M_2=400$ show how photon-rich Higgsinos (large $M_1$, small $\mu$) can be excluded.
Finally, with $M_2=250$, we see that $\gamma\gamma+\met$ can also exclude mixed wino-Higgsino NLSPs (e.g.\ $M_2=250$ and $\mu\approx 200$ with $M_1\to\infty$), where the branching fraction to photons is due to partly to the wino content, as in \ref{eqn:winodecay}. In the last two examples, note the difference between the left and right plots in fig.\ \ref{fig:exclggmetgen} -- in the latter, the bounds are weaker because the branching fraction to photons via the Higgsino is $\tan\beta$ suppressed. 

For wino co-NLSPs and heavier Higgsino NLSPs, the $\gamma\gamma+\met$ channel is not useful. For Higgsino NLSPs, the branching fraction to photons is a result of phase-space suppression in the other channels, so making the NLSP heavier turns off the rate in $\gamma\gamma+\met$. For wino co-NLSPs,   the $\gamma\gamma+\met$ signature can only come from direct production of $\tilde \chi^0_1 \tilde \chi^0_1$. However, as noted above, the cross section for this is negligible in the wino co-NLSP limit, since only Higgsino neutralinos can be directly produced.

Finally, let us specialize to the case of bino NLSPs and discuss the bound from $\gamma\gamma+\met$ here. As discussed in the previous section, in the case of a bino NLSP, the production cross section is determined by the mass of the heavier chargino states and is essentially independent of the bino mass. Here we consider the case that these heavier states are winos (i.e. $|M_2| \ll |\mu|$). Comparing (\ref{eqn:ggmetbound}) and the plot of the -ino cross section in fig.\ \ref{fig:genxsecs}, we see that the lower bound on the chargino mass ($M_2$) is\footnote{This result is valid in the large $\mu$ limit, and is the weakest possible bound in the scenario $\mu \gg M_2 > M_1$. The stronger bound of 300 GeV quoted by CDF relies on a particular SPS benchmark model point.}
\beq
\label{eqn:ggmetbino}
 m_{\chi_1^\pm} > 270\,\,{\rm GeV} 
 \eeq
In 10 fb$^{-1}$, the projected bound on the chargino mass is:
\beq
\label{eqn:ggmetbinoproj}
 m_{\chi_1^\pm} > 300\,\,{\rm GeV} \qquad (10\,\,{\rm fb}^{-1}\,\,{\rm projected})
 \eeq

\begin{figure}[!t]
\begin{center}
\includegraphics[width=0.35\textwidth]{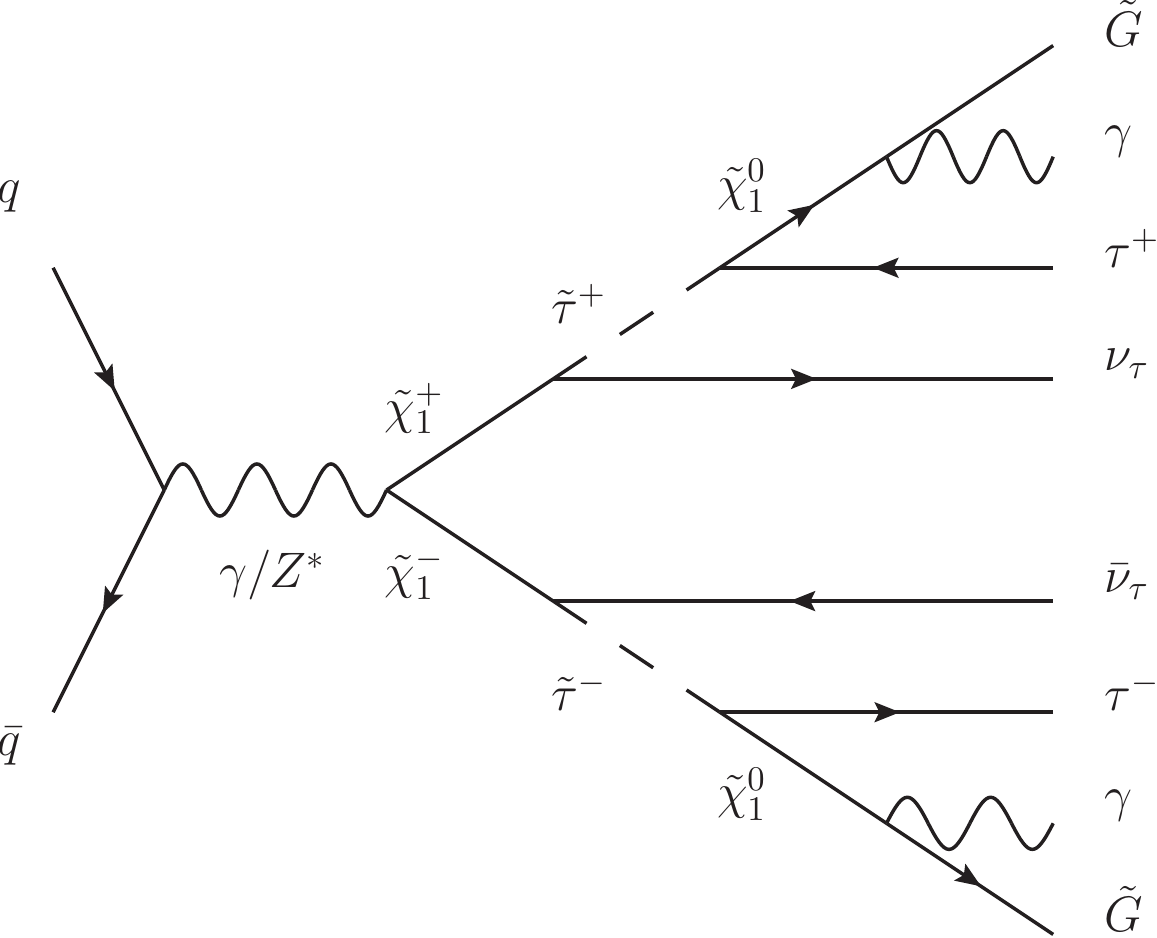}~~~~~\includegraphics[width=0.5\textwidth]{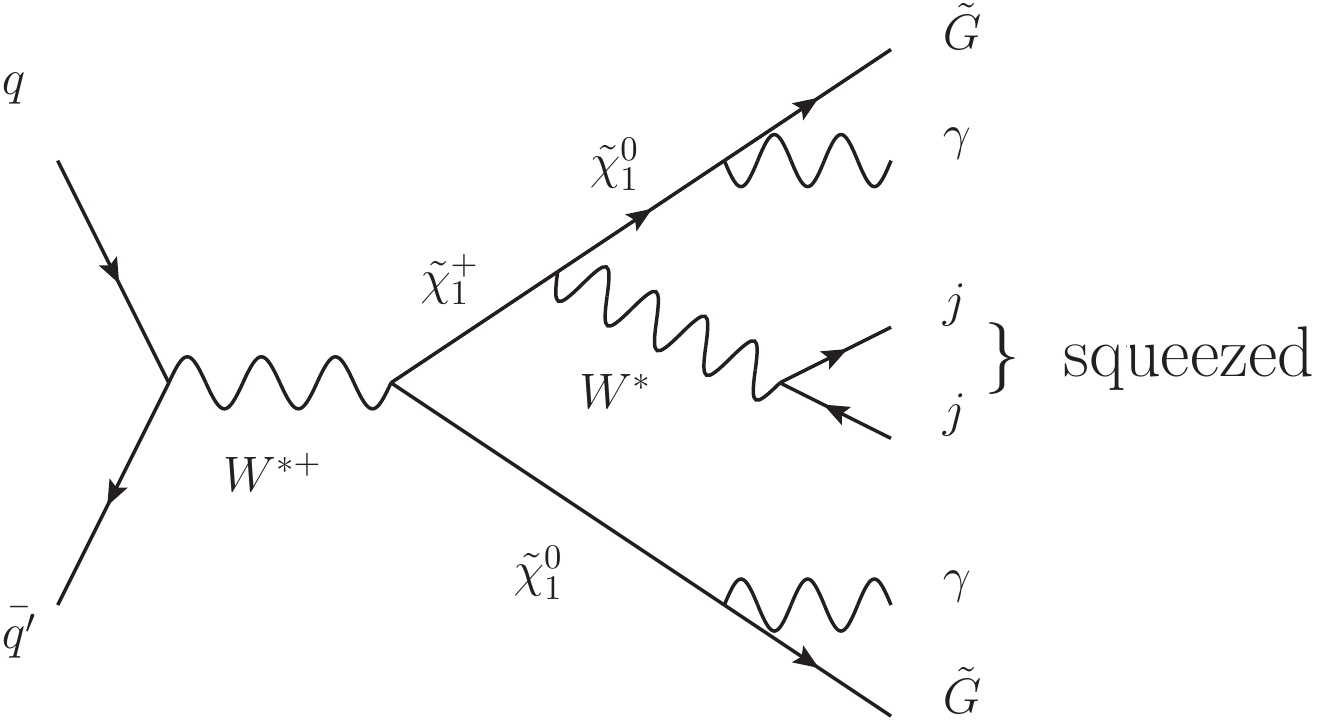}
\end{center}
\caption{Feynman diagrams for two-photon processes arising from neutralino and chargino production. At left, the typical process in MGM, where $\tilde{\chi}^\pm_1$ are mostly wino and decay through sleptons to the mostly-bino $\tilde{\chi}^0_1$. The final state includes energetic tau leptons. At right, a typical process with mostly-Higgsino NLSPs, which are produced directly. The small splitting between $\tilde{\chi}^\pm_1$ and $\tilde{\chi}^0_1$ leads to a three-body decay through off-shell $W$ with very little phase space, so there are relatively soft leptons or jets in the final state.} \label{fig:feyndia}
\end{figure}

We should point out that the kinematics and event topologies can be very different across parameter space. So if an excess is observed in this channel in future searches, such differences can be important in characterizing the physics. The main point is that for bino NLSPs, the rate is set by the masses of heavier winos, which can be far above the bino mass. Thus events will typically have energetic decay products coming from $\tilde W\to \tilde B + X$. For instance, in MGM, the decay chain can go through a slepton and one expects energetic leptons or taus in the events. However, for Higgsino NLSPs, the production cross section is determined mostly by the mass of the NLSP itself, with charged states split from neutral states by an amount of order 1-10 GeV. The decay products of heavier -inos down to the NLSP will be much softer, and the event could contain little additional activity beyond  $\gamma\gamma+\met$. Examples of the different decay chains are shown in fig.\ \ref{fig:feyndia}.

Various other Tevatron searches involving energetic photons and missing $E_T$ exist \citep{CDFgjmet, CDFgmet, D0gmet}. We have analyzed them in some detail; while some parts of parameter space can be excluded with these results, we find that the limit from $\gamma\gamma+\met$ is always much stronger, and so we will not discuss them in detail.

\section{Searches Relevant to wino co-NLSPs}
\label{sec:conlsp}
\setcounter{equation}{0} 

\subsection{Searches for $\gamma+W+\met$}

CDF has published a search for $\gamma+\ell+\met$ with 0.93 fb$^{-1}$ of data \citep{cdflgX}. They selected for at least one isolated photon and at least one isolated lepton ($e$ or $\mu$) with $p_T>25$ GeV and $|\eta|<1$. They also required $\met>25$ GeV. They found 163 events with an expected background of $150.6\pm13.0$. This null result sets a 95\% confidence limit on the cross section times branching fraction for general neutralino NLSPs:
\beq
\sigma\times \Br\times\varepsilon < 40\,\,{\rm fb}
\eeq
With 10 fb$^{-1}$, the projected bound is
\beq
\sigma\times \Br\times\varepsilon < 8\,\,{\rm fb}
\eeq
These bounds can be used to constrain wino co-NLSPs, in which one side is a chargino that decays promptly to $W+\tilde G$, and the other side is a neutralino that decays to $\gamma+\tilde G$. 

\begin{figure}[!t]
\begin{center}
\includegraphics[width=\textwidth]{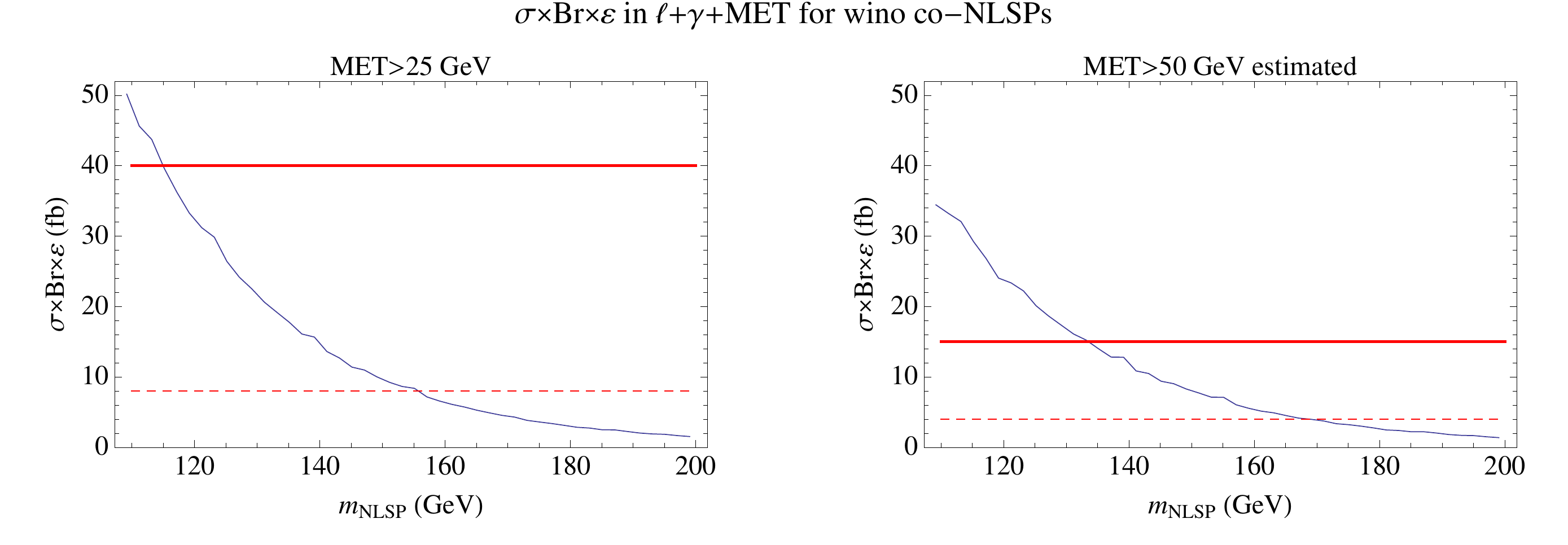}
\end{center}
\caption{Plots of $\sigma\times \Br\times\varepsilon$ (i.e.\ the number of events per fb$^{-1}$) in the $\ell+\gamma+\met$ channel coming from wino co-NLSPs. The left plot shows the number with $\met>25$ GeV, corresponding to the cut used in the CDF study. The right plot shows the number with $\met>50$ GeV. The solid lines indicate the 95\% confidence limit calculated from the CDF study, while the dashed lines indicate the projected bound in 10 fb$^{-1}$.} 
\label{fig:winoxsec}
\end{figure}

Shown in fig.\ \ref{fig:winoxsec} is a plot of $\sigma\times \Br\times\varepsilon$ in the $\gamma+\ell+\met$ channel coming from wino co-NLSPs. Here the signal acceptance $\varepsilon$ was computed from PGS simulations, as described in section \ref{subsec:pgs}.
From the figure, we see that the existing CDF search already bounds wino co-NLSPs,
\beq
 m_{NLSP} > 115\,\,{\rm GeV} 
  \eeq
With 10 fb$^{-1}$, the bound potentially becomes
\beq
 m_{NLSP} > 155\,\,{\rm GeV} \qquad ({\rm 10\,\,fb}^{-1}\,\,{\rm projected})
  \eeq

With simple modifications to the existing analysis, we find that the sensitivity can be extended significantly. In particular, we find that the $\met$ coming from SM $WW$ production is generally much less than the $\met$ in the signal. The second plot in fig.\ \ref{fig:winoxsec} shows our estimate of the bounds if a $\met>50$ GeV cut is applied: 
\beq
\label{eqn:winobound} m_{NLSP} & > & 135\,\,{\rm GeV}\qquad (\met>50\,\,{\rm GeV})\\ 
\label{eqn:winoprojbound} m_{NLSP} & > & 170\,\,{\rm GeV}\qquad (\met>50\,\,{\rm GeV},\,\,\,{\rm 10\,\,fb}^{-1}\,\,{\rm projected})
\eeq
The backgrounds were derived by using the $\met$ distribution shown in fig.\ 4 of \citep{cdflgX}. We see that a harder $\met$ cut improves the existing bound and the projected bound by 15-20 GeV. It is possible that an even harder  $\met$ cut would improve things further, but for that we would need a more accurate determination of the backgrounds at very large $\met$ than is available in \citep{cdflgX}. Also, it would be interesting to investigate cutting on other kinematic variables such as $H_T$ and $M_T$.\footnote{The Standard Model $W\gamma$ process has a large charge asymmetry, which may also be useful in distinguishing signal from background in this channel. We thank Yuri Gershtein for this observation.}

Given the significant statistical improvement that a harder $\met$ cut and $10\times$ more data will bring to the existing search, it is also worth mentioning the discovery potential in this channel. We estimate that a 5$\sigma$ discovery is possible up to $m_{NLSP}\approx 140$ GeV, and 3$\sigma$ ``evidence" is possible up to $m_{NLSP}\approx 160$ GeV.

Finally, let us briefly discuss the possibility of using $\gamma+{\rm jets}+\met$ to constrain the wino NLSP scenario. Here obviously one is substantially increasing the signal rate through the higher branching fraction of $W\to{\rm jets}$, but one is also increasing the SM backgrounds by introducing hadronic activity into the event. There do not appear to be any public Tevatron searches in this channel; the closest seems to be \citep{CDFgjmet}, which unfortunately searched specifically for events with displaced photons. Also, preliminary results on $\gamma+{\rm jets}(+\met)$ were presented in \citep{Dittmann:2008ui}; however, at present this analysis has no $\met$ cut, and therefore the backgrounds are simply too high for it to be useful. Without a more detailed analysis, it remains to be seen whether signal or background will win out in the end.

\subsection{Other searches}

There have been several Tevatron analyses of the $W^+(\ell^+\nu) + W^-(\ell^-\nu)$ final state, including a CDF measurement of the $W^+W^-$ cross section with 3.6 fb$^{-1}$  of data \citep{CDFWWllmet}, and a \dzero measurement of the same cross section with 1 fb$^{-1}$ of data \citep{D0WWll}. We find that neither search constrains wino NLSPs even with 10 fb$^{-1}$ of integrated luminosity. So this channel does much worse than the $\gamma+W(\ell\nu)$ channel considered in the previous subsection. The reasons appear to be: (1) the relative signal suppression due to the extra factor of $\Br(W\to\ell\nu)$ and the fact that $\sigma(p\bar p\to \chi_1^+\chi_1^-)\approx {1\over2}\sigma(p\bar p\to \chi_1^+\chi_1^0)$ in the range of NLSP masses of interest; and (2) the high background rate $\sim \CO(10^2)\,\, {\rm fb}$ from a variety of sources. In particular, as discussed in section \ref{sec:searches}, SM $WW$ production has a rather high cross section relative to the signal, and it is very difficult to eliminate this source of background. We have investigated a number of simple kinematic variables, such as $\met$, and have not found one that can significantly reduce SM $WW$ while keeping enough signal. Perhaps a more sophisticated kinematic quantity such as $M_{T,2}$ can help \citep{LesterSummers,BarrLesterStephens}.

We have similarly examined the CDF analysis in the $Z(e^+e^-)+W({\rm jets})+\met$ final state using 2.68 fb$^{-1}$ of data \citep{CDFWZmet}. Here the main analysis cuts were: central, high-$p_T$, opposite-sign di-electron pair forming a $Z$; a dijet pair forming a $W$; and $\met>$40, 50 or 60 GeV.
With these three different $\met$ cuts, CDF observed 7, 2, and 1 event with 6.41, 3.76 and 2.02 expected, respectively. Simulating the signal, we find that this analysis does not constrain wino NLSPs even with 10 fb$^{-1}$ of data. The problem again seems to be the low signal rate; as discussed in section \ref{sec:searches}, the wino NLSP cross section is below 4 fb everywhere, and the signal acceptance is at most 30\% just from geometric considerations.

\section{Searches Relevant to $Z$-rich Higgsino NLSPs}
\label{sec:zsearches}
\setcounter{equation}{0} 

\subsection{Searches for $Z(\ell^+\ell^-)+\met+X$}

Results in this channel include a CDF search for $Z(\ell^+ \ell^-) + \met (+X)$ \citep{CDFZXY}, a \dzero measurement of the SM $Z(\ell^+ \ell^-) + Z(\nu\nu)$ cross section \citep{D0ZZllnunu}, and a CDF search for $Z(e^+e^-) + W({\rm jets})+\met$ \citep{CDFWZmet}. The dominant SM backgrounds for all these Tevatron searches consist of $Z+{\rm jets}$, $W+{\rm jets}$, $t\bar t$, and dibosons. 

We find that none of the existing studies currently rule out any part of the general parameter space. However, with 10 fb$^{-1}$ of data, all of these searches have the potential to exclude regions of Higgsino and wino-like parameter space, and some, with cuts optimized for our scenario, even have the possibility for genuine 5$\sigma$ discovery. In particular,  the search for $Z(\ell^+ \ell^-) + \met +H_T$ in  \citep{CDFZXY} appears to be the most effective. So let us now briefly describe the search and its implications for general neutralino NLSPs.

The requirements were pairs of central, opposite-sign electrons or muons (one with $p_T > 20$ GeV, one with $p_T > 12$ GeV) reconstructing a $Z$ within the window 66 GeV $< M_{inv} <$ 116 GeV. Further requirements were $\met > 25$ GeV with 3-sigma significance (i.e. MetSig $> 3$, as defined in \citep{CDFZXY}) and $H_T > 300$ GeV. With these cuts, CDF observed 6 $e^+e^-$ events with $6.6\pm1.0$ expected, and 6 $\mu^+\mu^-$ events with $3.5\pm0.5$ expected, with an integrated luminosity of 0.94 fb$^{-1}$.

\begin{figure}[!t]
\begin{center}
\includegraphics[width=\textwidth]{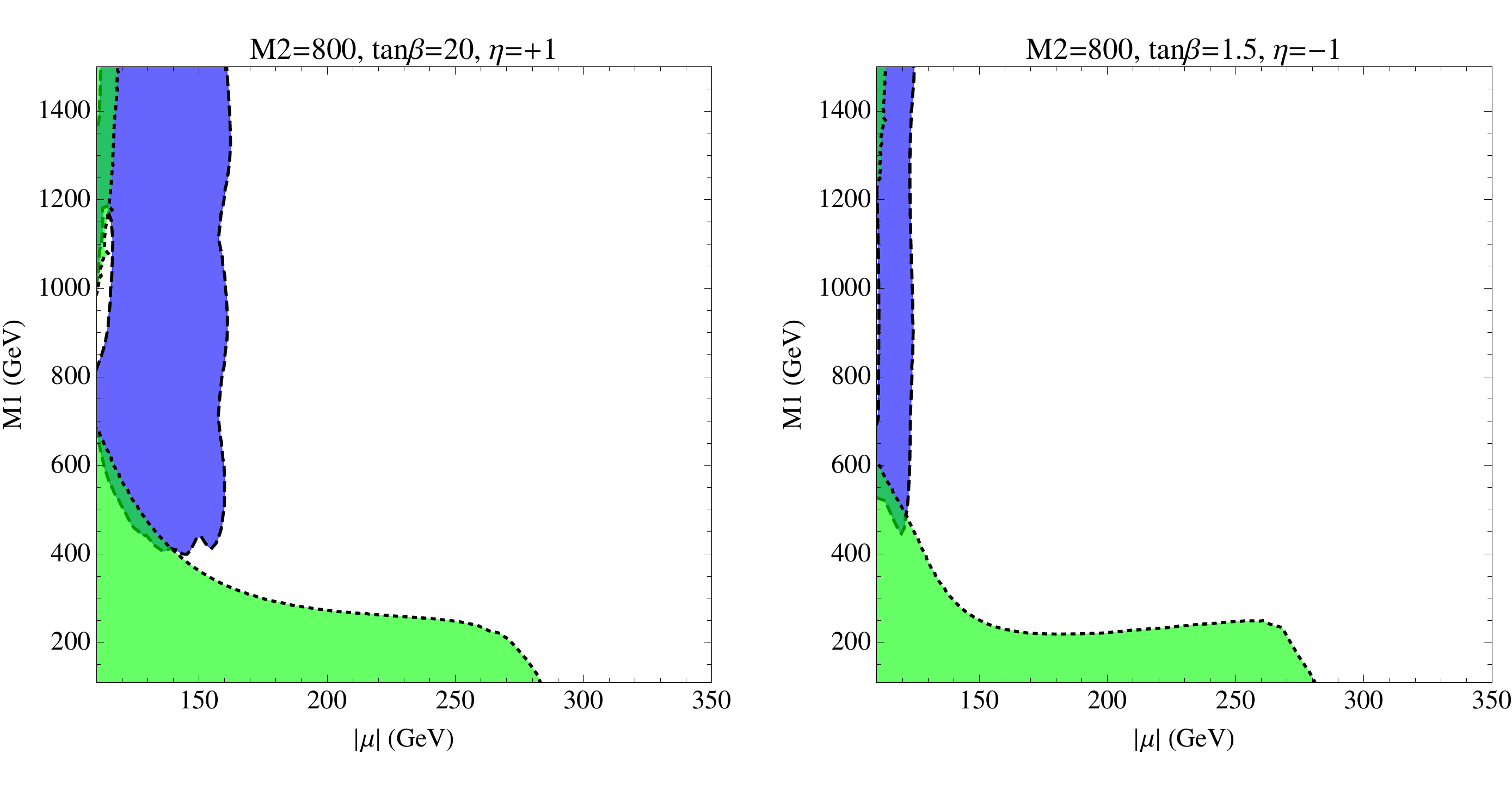}
\end{center}
\caption{Various examples of projected 95\% exclusion contours (blue, dashed)  in the $(\mu, M_1)$ plane, based on the CDF search for anomalous $Z(\ell^+\ell^-)+H_T+\met$ production. The contours are derived assuming 10 fb$^{-1}$ of data. Also shown are the projected exclusion contours (green, dotted) for the $\gamma\gamma+\met$ search, for the same choices of the parameters. } 
\label{fig:exclZHTMETgen}
\end{figure}

The projected 10 fb$^{-1}$ exclusion contours in the $(\mu,\,M_1)$ plane are summarized by the plots in fig.\ \ref{fig:exclZHTMETgen}. For comparison, the $\gamma\gamma+\met$ contours for the same choices of the parameters are also shown. We see that the two channels are mostly complementary in their coverage of the parameter space. The exclusion contours from the $Z+H_T+\met$ search are largely independent of the other parameters ($M_2$, $\tan\beta$, $\eta$), and the typical situation is characterized by the left plot. The one exception, shown in the right plot, is when $\tan\beta$ is small and $\eta=-1$. As discussed in  section \ref{subsec:HiggsinoNLSP}, in this case the branching fraction to Higgses takes over at larger NLSP mass, and this weakens the $Z+H_T+\met$ sensitivity.

The pure Higgsino NLSP limit corresponds to $M_1\to \infty$ in the plots in fig.\ \ref{fig:exclZHTMETgen}. We see that in this limit, the projected bound is
\beq
m_{NLSP} > 150\,\,{\rm GeV}\qquad (10\,\,{\rm fb}^{-1}\,\,{\rm projected})
\eeq
essentially independent of the other parameters, except again when $\eta=-1$ and $\tan\beta\to 1$. In this case, we estimate that the  $Z(\ell^+\ell^-)+H_T+\met$ search should be able to cover the entire $Z$-rich range of NLSP mass. For instance, if $\eta=-1$ and $\tan\beta=1.5$ (the case shown in figs.\ \ref{fig:HiggsinoBR} and \ref{fig:exclZHTMETgen}), we find that the projected bound is $m_{NLSP}>130$ GeV. For larger values of $m_{NLSP}$ the branching fraction to Higgses takes over.

\subsection{Suggestions for Optimizing $Z+\met+X$ Searches}

We have seen in the previous subsection that with enough integrated luminosity, existing searches at the Tevatron can potentially exclude Higgsino NLSPs with mass up to $\approx$ 150 GeV. However, these searches are not targeted specifically to the Higgsino NLSP scenario, and as we will discuss in this subsection,  some simple optimizing of the cuts can easily extend the reach by $\sim 20$ GeV.

\begin{figure}[!t]
\begin{center}
\includegraphics[width=\textwidth]{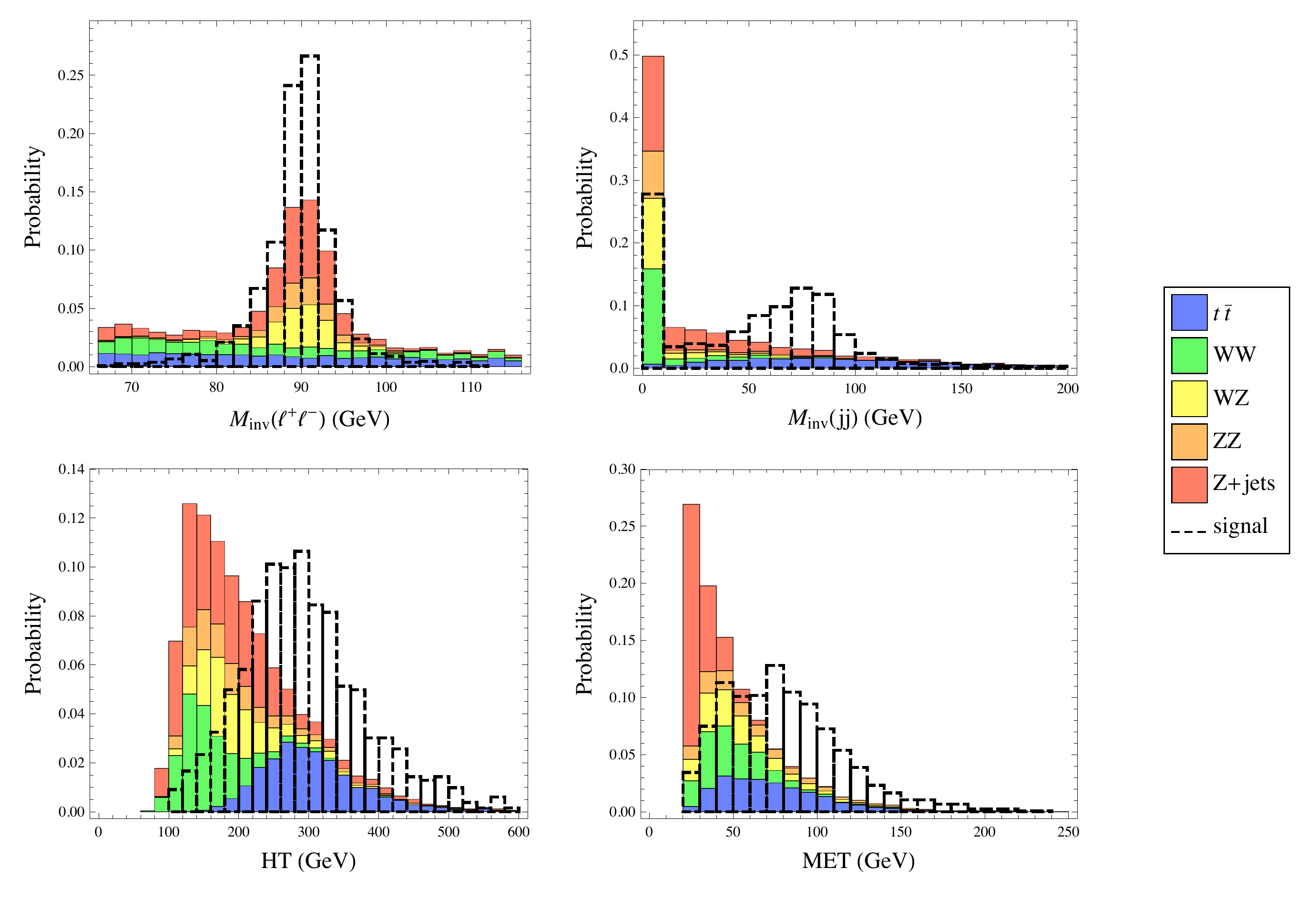}
\end{center}
\caption{Comparison of distributions for a Higgsino NLSP signal point ($\mu=170$ GeV, $M_1$, $M_2\to\infty$, $\tan\beta$ = 20) and backgrounds. Signal and total background are independently normalized to unit area. In every case we plot only for those events passing the cuts in the $Z(e^+e^-)+\met$ analysis of \citep{CDFZXY}, i.e.\ $\met > 25$ GeV, MetSig$>3$, and a central $Z$ decaying to $e^+e^-$. For the $M_{inv}(jj)$ distribution, the invariant mass of the hardest two jets is plotted, and the zero bin is populated by events with less than two jets.} 
\label{fig:proposedhistos}
\end{figure}

In fig.\ \ref{fig:proposedhistos}, we compare a few kinematic distributions for the main SM backgrounds and a particular Higgsino NLSP signal point. These distributions were chosen to highlight the differences between background and signal. And apart from $M_{inv}(jj)$, they also happen to be the kinematic variables used in the CDF analysis. While we have not performed a comprehensive analysis, other distributions that we have examined, such as $N_{jets}$, $N_{b-jets}$, $\cos\Delta\phi(Z,\met)$, etc., do not seem to separate background and signal as cleanly.

From the figure, we see that simple modifications of the CDF cuts can potentially improve the sensitivity to our signal hypothesis. For instance,  tightening the $Z$ mass window cut may be useful. This was set at 66 GeV $< M_{\ell^+ \ell^-} <$ 116 GeV in the CDF analysis \citep{CDFZXY}, so as to include the entire range in which the $Z$ peak in Drell-Yan rises above background. We see from the figure that a narrower range, such as $M_{\ell^+\ell^-}\in(80,\,100)$ GeV, will still keep most of our signal, while cutting down on backgrounds without real $Z$s ($W^+W^-$ and $t\bar{t}$) significantly.

The $H_T$ distribution in  fig.\ \ref{fig:proposedhistos} shows that the $H_T>300$ GeV cut used in the CDF analysis tends to eliminate too much of the signal. This is simply because the $H_T$ in the signal is coming entirely from the pair-produced -ino decay products, and the -ino masses considered here are in the $\sim 100-200$ GeV range. According to the figure, a lower $H_T$ threshold, such as $H_T>200$ GeV, would boost the signal acceptance while still eliminating most of the backgrounds except $t\bar t$.

Finally, the $\met$ distribution in fig.\ \ref{fig:proposedhistos}\ shows that a harder $\met$ cut, such as $\met>40$ GeV, can improve $S/\sqrt{B}$ significantly.\footnote{A brief comment on our background simulations: with a harder $\met$ cut, we are potentially underestimating $Z+{\rm jets}$, which comprises $\sim 20\%$ of the background after the cuts described here. The reason is that PGS does not accurately reproduce the tails of the $Z+{\rm jets}$ $\met$ distribution, since it is mostly coming from mismeasurement.  However, our use of the MetSig variable defined in \citep{CDFZXY} should help, since this involves detailed information about jet energy resolution. Indeed, our simulations are able to reproduce to 10-20\% accuracy all of the analyses in \citep{CDFZXY} which involve MetSig. So hopefully this means that the background estimates used in the modest extrapolation described here are not too far off.}

Motivated by these considerations, we find by optimizing the analysis of $Z+\met+X$ along these lines, the reach can be improved to
\beq
m_{NLSP} \gtrsim 170\,\,{\rm GeV}\qquad (10\,\,{\rm fb}^{-1}\,\,{\rm projected})
\eeq
for the $Z$-rich Higgsino NLSP case. Also, we find that $5\sigma$ discovery is possible up to $m_{NLSP}\approx 135$ GeV and $3\sigma$ evidence up to $m_{NLSP}\approx 160$ GeV. The background and signal ($m_{NLSP}=170$, $\eta=+1$ and $\tan\beta=20$) counts are shown in table \ref{tab:Zmetproposed} for the following set of cuts: 80 GeV $< M_{\ell^+ \ell^-} <$ 100 GeV, $H_T>200$ GeV, $\met>40$ GeV, MetSig$>3$.

\begin{ctable}[
caption = Rates for a proposed $Z+H_T+\met$ search,
label = tab:Zmetproposed,
pos = t]{|c|c|c|c|c|}
{
}
{
\hline
$N_{bg}(e^+e^-)$ (fb) & $N_{bg}(\mu^+\mu^-)$ (fb) & $N_{sig}(e^+e^-)$ (fb)& $N_{sig}(\mu^+\mu^-)$ (fb)\\
\hline
\hline
7.4 & 5.8 & 1.4 & 1.1\\
\hline
}
\end{ctable}

Additional cuts could potentially play a role, but we have not investigated them in detail. For instance, $t\bar{t}$ background can be reduced by a cut on the second-hardest jet in the event. Mass window cuts could be used to attempt to isolate a second $Z$ boson decaying to jets (as indicated in the $M_{inv}(jj)$ distribution shown in fig. \ref{fig:proposedhistos}), or a Higgs decaying to $b\bar{b}$ (perhaps with $b$-tagging, as well). As we discussed in the previous section, the splitting between charged and neutral Higgsinos can be quite small, and will lead to fairly soft jets or leptons. So selecting events with a soft, isolated lepton from the decay $\tilde\chi^+_1 \to \ell^+ \nu \tilde\chi^0_1$ could help to purify the signal. Needless to say, if an excess is observed in this channel, any of these features, even if not useful for obtaining good results in a counting experiment, could help in understanding whether or not an NLSP explanation fits the data. Further improvement might come from considering $Z \to \tau^+ \tau^-$ final states, but due to the neutrinos in the $\tau$ decays, such events would be unlikely to pass the tight $Z$ mass-window cut, and one would have to reconsider how to reject backgrounds.

Finally, we would like to emphasize that with very simple, naive cuts alone we have seen that the sensitivity in this channel can be extended by $\sim 20$ GeV. After these cuts, we are still finding $\sim 25$ signal events in 10 fb$^{-1}$. This strongly suggests that with smarter cuts and more sophisticated experimental methods (e.g.\ neural nets), the sensitivity could be extended even further.

\subsection{Searches For $Z(\ell^+\ell^-)+Z(\ell'^+\ell'^-)$}
\label{sec:ZllZll}
\setcounter{equation}{0} 

Here we discuss Tevatron measurements of the SM $ZZ$ cross section using the  $Z(\ell^+\ell^-)+Z(\ell'^+\ell'^-)$ final state. The most recent published analyses are one from CDF with 1.9 fb$^{-1}$ of data \citep{CDFZZ4lprl} and one from \dzero with 1.7 fb$^{-1}$ of data \citep{D0ZZ4lprl}. In addition, there is a very recent public note of a CDF analysis with 4.8 fb$^{-1}$ of data \citep{CDFZZ4l}. All of these analyses require two opposite sign dilepton pairs satisfying various kinematic and isolation requirements.

We have chosen to reproduce the published 1.9 fb$^{-1}$ CDF analysis for simplicity. They observe 2 events in  1.9 fb$^{-1}$, and the SM prediction (coming almost entirely from $ZZ$ production) is 2.004$^{+0.016}_{-0.015}\pm0.210$. 

Simulating the signal, we find that the existing CDF analysis has no sensitivity in the general neutralino NLSP parameter space. Furthermore, even in 10 fb$^{-1}$ the projected 95\% exclusion contours are negligible, and are completely covered by $\gamma\gamma+\met$ and $Z+H_T+\met$. The reason for this is simply because the overall rate is suppressed by two factors of the small branching fraction of $Z\to \ell^+\ell^-$, and so $S/\sqrt{B}\lesssim 1$ everywhere in the parameter space.

\begin{figure}[t!]
\begin{center}
\includegraphics[width=0.65\textwidth]{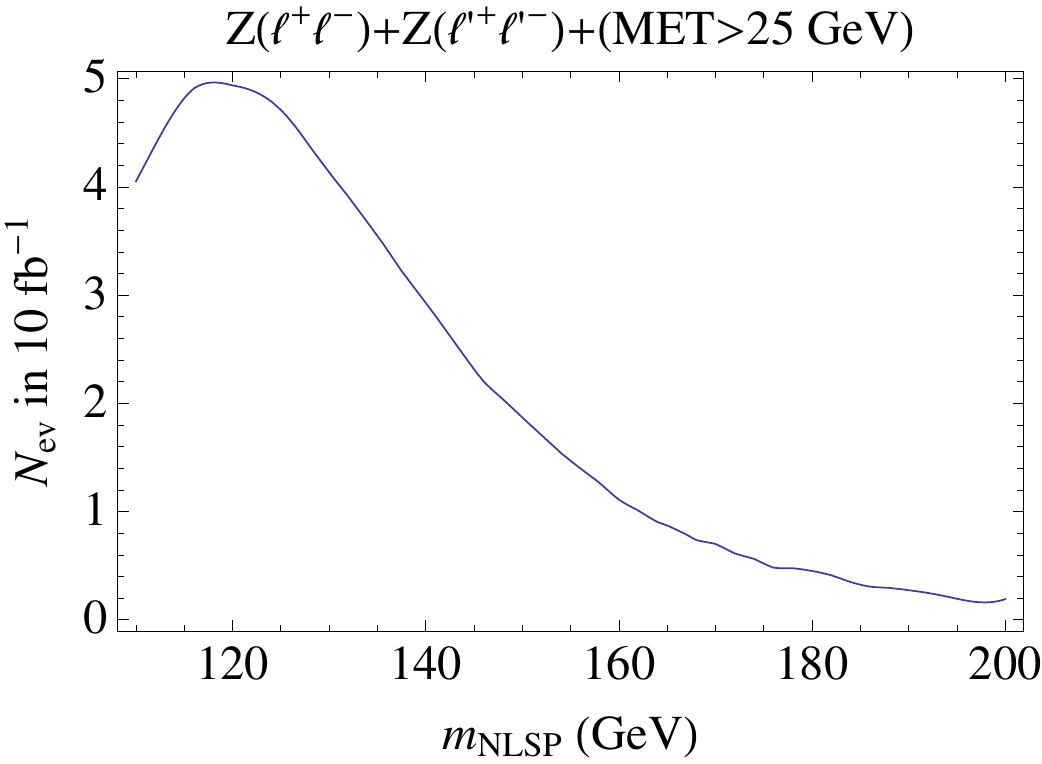}
\end{center}
\caption{The projected number of Higgsino NLSP events in 10 fb$^{-1}$ in a hypothetical search for $Z(\ell^+\ell^-)+Z(\ell'^+\ell'^-)+\met$, with $\met>25$ GeV.} 
\label{fig:ZllZllmet}
\end{figure}

However, things could be improved with a $\met$ cut, since the background (SM $ZZ$ production) should have very little intrinsic $\met$. We do not attempt to simulate fake $\met$ from SM $ZZ$ production, we simply assume that a reasonably sized $\met$ cut, e.g.\ $\met>25$ GeV, should reduce the background to $N_b\ll 1$.  Then we would like to have at least a few events to claim a discovery.  A plot of the estimated number of Higgsino NLSP events in 10 fb$^{-1}$ in this hypothetical search is shown in fig.\ \ref{fig:ZllZllmet}. Here $M_1$, $M_2$ are taken to be very large, $\mu>0$, and $\tan\beta=20$. We see that, assuming the SM background is very small, we could conceivably have a ``discovery" for $m_{NLSP}$ up to $\sim 140$ GeV. Of course, given the low statistics, a true estimate of the discovery potential would require a much more accurate estimate of the small SM background.
We should also emphasize that the low statistics highly motivates a D0/CDF combined analysis in this channel.

\section{Searches Relevant to $h$-rich Higgsino NLSP}
\label{sec:bjets}
\setcounter{equation}{0}

Finally, we come to the most difficult scenario to analyze, in which the NLSP decays dominantly to $h + \tilde G$. It occurs in a relatively small part of parameter space (Higgsino NLSP with $\eta =-1$ and $\tan \beta \approx 1$), but is interesting, especially for the prospect of observing the Higgs at the same time as physics beyond the Standard Model. Because one can have Higgs bosons on both sides of the event, each decaying to $b\bar{b}$, the signals can involve anywhere from two to four $b$-tagged jets along with missing $E_T$. 

Before considering the multi-$b$ channel in detail, we note that the channel $\gamma+b+j+\met$ is potentially useful in a small slice of parameter space where $\tan \beta \approx 1$ and $M_1 \approx -\frac{4}{3} \mu$. (In general, one does not have $Br(\gamma) \approx Br(h) \sim O(1)$; see the lower right-hand plot of Figure \ref{fig:HiggsinoBR}.) Early in Run II, a Tevatron SUSY Working Group considered this channel in some detail \citep{WorkingGroup}, choosing a ``Higgsino Model Line I" in this slice. Recent observations in this channel show a rate of $\approx$15 fb for events with $\met > 50$ GeV, consistent with the backgrounds, which are fake-dominated, with large systematic uncertainties \citep{CDFgbjmet}. The early working group study relied on lower background estimates extrapolated from the small Run I dataset. Hence, we do not consider $\gamma+b+j+\met$ a very promising channel. We will confine our attention to studies involving multiple $b$ jets plus $\met$ in the remainder of this section.

Backgrounds in the multi-$b$+$\met$ channel are large and uncertain. We can get a sense of how feasible a search is by looking for similar existing new physics searches. Searches for more than 2 $b$-tagged jets are motivated by the prospect of MSSM Higgs bosons \citep{CDFh3b, D0multib}. Most recently, such a study from \dzero \citep{D0multib} required 3 jets with tight $b$-tags in events with up to 5 jets and loose acceptance cuts. In a sample of 1 fb$^{-1}$, they found 3224 3-jet events, 2503 4-jet events, and 704 5-jet events. These samples are much too large for our signal to show up in a simple counting experiment. Furthermore, \dzero set limits using only the shapes, not counts, due to uncertain backgrounds. The $M(bb)$ distributions in the data peak just above 100 GeV, so a cut in this variable will not help us. A hard $\met$ cut is more promising, but we do not have access to the $\met$ distribution in the data to read off a reliable count.

We learn more from existing searches for $b\bar{b}+\met$. A CDF search for ZH/WH production \citep{CDFbbmet} and a \dzero search for leptoquarks or sbottoms \citep{D0bbmet} each look for two $b$-jets and $\met$, but due to a veto on events with 4 or more jets, they fail to constrain the Higgsino scenario. However, similar searches are potentially constraining, as we will see shortly. More interesting is a CDF search for gluino-mediated sbottom production using 2.5 fb$^{-1}$ \citep{CDF4bmet}, which looks for $\met > 70$ GeV and two $b$-tagged jets with some basic kinematic cuts. In a sample that further rejects leptons and requires $\met$ to be far from jets in $\phi$, they observe 451 events with 506 $\pm$ 144 expected. Note that the rates and the uncertainties are large, even with a strict $\met$ cut. Top quark and heavy-flavor multijet backgrounds are dominant.

Owing to their high multiplicity and their all-hadronic, heavy-flavor content, we do not attempt to simulate the multi-$b$+$\met$ backgrounds in detail, as in the previous sections. However, we can give some preliminary remarks. We have simulated $t\bar{t}$ and diboson ($WW$, $WZ$, and $ZZ$) samples in Pythia \citep{Pythia}, and $Zb\bar{b}$, $Wb\bar{b}$, and $b\bar{b}b\bar{b}$ samples in MadGraph \citep{MadGraph}. These background samples were processed with PGS, which includes a parametrization of CDF's SecVtx $b$-tagging efficiency as a function of $E_T$ and $\eta$. For a messy hadronic channel like this one, backgrounds should be simulated with matching, and run through the collaboration's full detector simulation, to obtain realistic estimates. Nonetheless, we offer a rough guide to how to begin to design a search.

\begin{figure}[t]
\begin{center}
\includegraphics[width=0.435\textwidth]{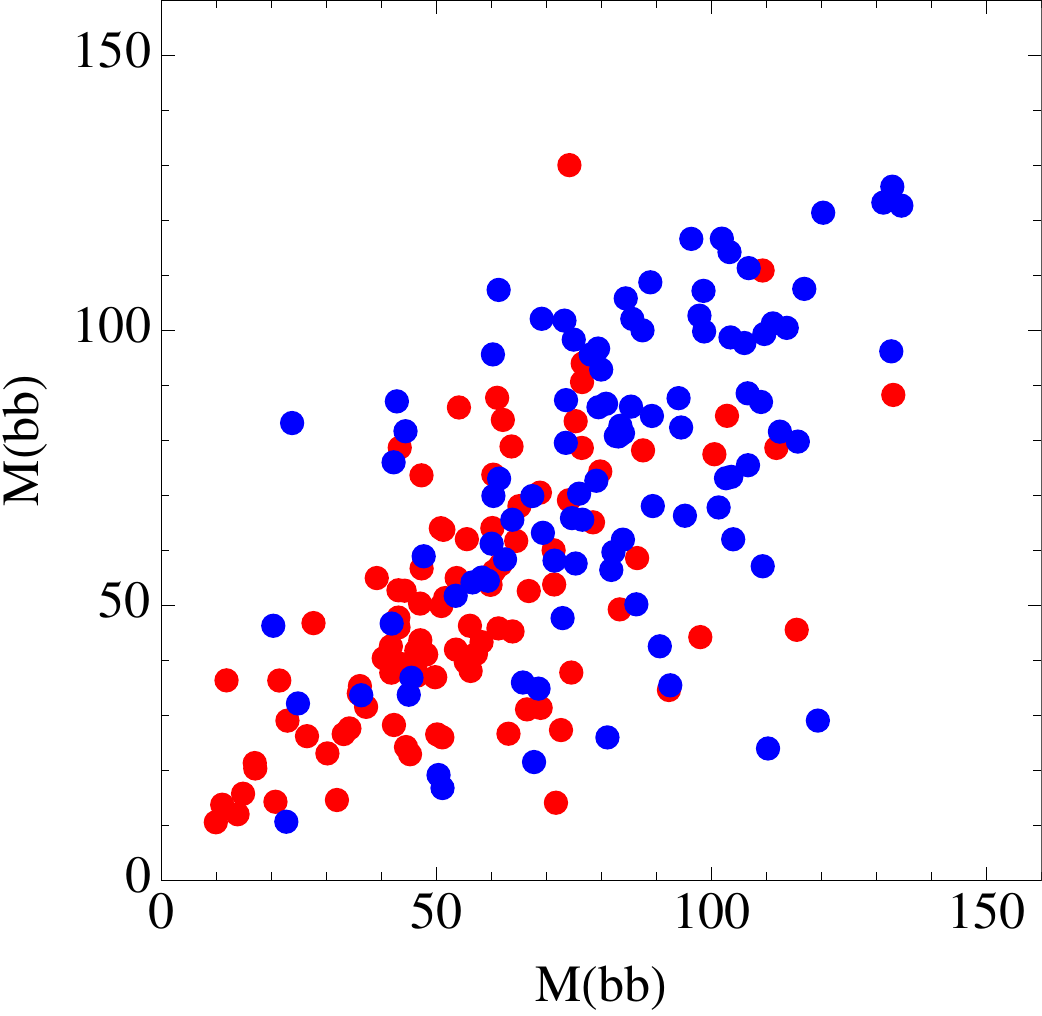}~~\includegraphics[width=0.41\textwidth]{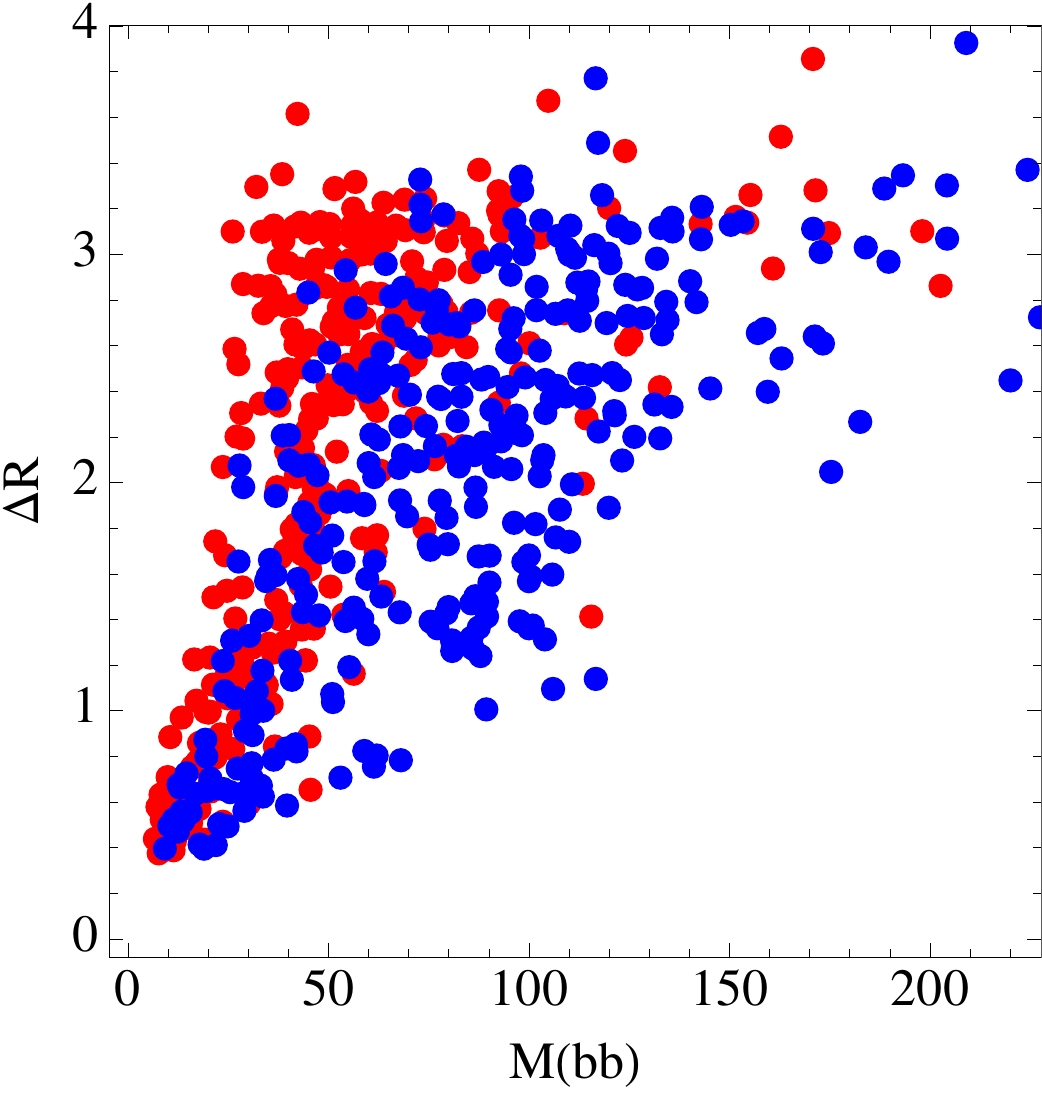}
\end{center}
\caption{At left: nearest pairs of $M(bb)$ in partitions of 4 b-tagged jets, in 100 signal events (blue) and 100 $b\bar{b}b\bar{b}$ background events (red). At right: all combinatoric possibilities for invariant mass and $\Delta R$ between pairs of b-tagged jets in 50 signal events with 4 b-tags (blue) and 50 $b\bar{b}b\bar{b}$ background events with 4 b-tags (red). The signal point is $\mu = -140$ GeV, $M_1 = 510$ GeV, $\tan\beta = 1.5$, $M_2 = 800$ GeV.}
\label{fig:BBvsBB}
\end{figure}

We first examine a set of cuts that asks for 50 GeV of $\met$, at least two $b$-tagged jets, and no central electrons or muons. The rate is still very large (as suggested by the results of Ref. \citep{CDF4bmet}), so we further require that at least one pair of $b$-tagged jets has an invariant mass satisfying 60 GeV $< M(b_1,b_2) < 200$ GeV, with $\Delta R(b_1,b_2) < 2.5$. The latter requirement is imposed because $b$-jet pairs in the background samples are unlikely to have large invariant mass if they are not back-to-back. (See the right-hand plot in Figure \ref{fig:BBvsBB}, which we will discuss in more detail shortly.) The results are tabulated in Column A of Table \ref{tab:bbmet}.

We next consider a second set of events with more $b$-tags: either three or more tight $b$-tags, or two tight plus at least one loose $b$-tags. This substantially reduces the $t\bar{t}$ background, which rarely has three $b$-tags unless one of the W decays gives a charm quark (which has $\approx 10\%$ probability of being $b$-tagged). Demanding more $b$-tags allows us to relax the $\met$ cut somewhat, so we only ask for 40 GeV of $\met$ in this sample. The results are in Column B of Table \ref{tab:bbmet}. 
Note that now the $b\bar{b}b\bar{b}$ sample is much more important than it was in the sample with two tags and a harder $\met$ cut.

\begin{ctable}[
caption = Multi-$b+\met$ rates in fb,
label = tab:bbmet,
pos = !t]{|c|c|c|c|}
{
\tnote[a]{Includes only simulated backgrounds -- not comprehensive.}
\tnote[b]{In all signal points listed, $M_1 = 500$ GeV, $M_2 = 800$ GeV, $\eta=-1$ and $\tan \beta = 1.5$.}
}
{
\hline
Sample & A. $bb$ + $\met>$50  & B. $bbb$ + $\met>$40  & C. $bbbb$+$\met>$30  \\
\hline
\hline
$t\bar{t}$ & 77.2 & 16.8 & 1.7\\
$Wb\bar{b}$ & 12.4 & 1.4 &  0.0 \\
$Zb\bar{b}$ & 6.1 & 0.8 & 0.1 \\
$b\bar{b}b\bar{b}$ & 4.8 & 9.1 & 1.6\\
Diboson & 2.7 & 0.2 &  0.0 \\
\hline
Total\tmark[a] & 103.2 & 28.3 & 3.4 \\
\hline
\hline
$m_{NLSP}$ = 140 GeV\tmark[b] & 5.2 & 3.1 &1.0\\
\hline
$m_{NLSP}$ = 160 GeV & 7.1 & 4.1  & 0.9\\
\hline
$m_{NLSP}$ = 180 GeV & 6.2 & 3.3 & 0.7\\
\hline
$m_{NLSP}$ = 200 GeV & 4.5 & 2.3 &  0.5\\
\hline
}
\end{ctable}

The final column in the table displays the rate for events with four $b$-tagged jets (up to two of which can have only loose tags), all with $p_T > 15$ GeV and $\met > 30$ GeV.  Unlike in columns A and B, no cut on the invariant mass of pairs has been imposed. The background rates, especially for $b\bar{b}b\bar{b}$, rise steeply when the $p_T$ cut on all jets, or the $\met$ cut, is loosened.

From the table, one can see that {\em if} these samples encompass all the important backgrounds, and {\em if} the $b$-tagging efficiencies are modeled well by PGS, and {\em if} the systematic errors are small enough to be neglected, then a counting experiment would be able to exclude some region from $\mu \approx -150$ GeV to $\mu \approx -180$ GeV at 95\% confidence. (We note that below $\mu\approx 150$ GeV, the Higgsino NLSP mostly decays to $Z$s, and this is covered by the analyses described in the previous section.) In practice, we expect that this is not the case. In particular, our samples overestimate the $t\bar{t}$ numbers reported in Ref. \citep{CDF4bmet} and underestimate the QCD heavy-flavor contribution (because we have not included $b\bar{b}jj$, etc.), and the estimates reported there carried large systematic uncertainties, which we have not even attempted to quantify for our samples. It is clear that a search will have to rely on the shapes of distributions, not just counting. From the results in Refs. \citep{CDFbbmet,D0bbmet,CDF4bmet}, we can be fairly confident that neural nets or other statistical techniques can eliminate much of the QCD multijet backgrounds with fake $\met$ or $\met$ arising from semileptonic $b$ or $c$ decays (which will tend to be aligned with the jet direction), so the results in the table may not be wildly optimistic. It appears that the largest background, especially if only two $b$-tags are demanded, will be $t\bar{t}$. Statistical analysis might be able to label events as ``top-like" and improve the discrimination. In the end, any limit set in this channel will probably rely on the fact that $t\bar{t}$ events have a broad distribution of $M(bb)$, which is more likely to be near $m_{h}$ in the signal.

One might ask whether the signal can be isolated just from invariant mass distributions. For instance, one could apply only a loose $\met$ cut, demand four $b$-tagged jets, and look for a pairing where $M(b_1 b_2) \approx M(b_3 b_4)$, with signal accumulating near the Higgs mass. However, we find that in the PGS output, such pairs do not cluster strongly near $(m_h, m_h)$, and offer little hope of finding a peak above the background (see the left-hand plot in Figure \ref{fig:BBvsBB}). Column  C of Table \ref{tab:bbmet} demonstrates that any technique that relies on tagging four $b$-jets will have limited signal rate, though with a $\met$ cut a signal-to-background ratio of order 1 may be achievable. One would want to rely on shape information, such as that illustrated in the right-hand plot of Figure \ref{fig:BBvsBB}, to gain confidence that a signal is real. This figure illustrates the correlation that was used in choosing cuts for columns A and B of Table \ref{tab:bbmet}. In the background events, pairs of jets with large invariant mass are rare, and typically such paired jets are nearly back-to-back ($\Delta R \gappeq \pi$). On the other hand, for the signal events, the two $b$ jets from a Higgs decay are typically not back-to-back (though they are well-separated). Notice that for $1 \lappeq \Delta R \lappeq 2$, $M(bb)$ in the signal cuts off at around $m_h$. Thus this shape offers some hope of discriminating signal from background and obtaining an estimate of $m_h$.

In practice, we expect that in an all-hadronic channel the understanding of the background shapes and normalization will have to be taken from data rather than Monte Carlo, and more sophisticated techniques such as neural nets or boosted decision trees could be useful. Further improvement could come from \dz, where $b$-tagging is efficient at larger values of $\eta$ than at CDF. (At CDF, soft lepton tagging in addition to SecVtx might also yield improvements beyond the efficiencies parametrized by PGS.) We expect that the general picture we have sketched --- using a $\met$ cut to reduce the overall size of the background, then studying the shape in the ($M(bb), \Delta R$) plane to identify the signal contribution --- could survive these complications and prove to be useful.

Our results are broadly consistent with those of the SUSY Working Group operating at the start of Run II \citep{WorkingGroup}. Their assumption (given the more limited Monte Carlo tools at the time) was that backgrounds were dominated by $t\bar{t}$. This is reasonable, so long as one only demands two $b$-tags. We have found that the multijet contributions that were not included in early studies are important when three or more $b$-tags are included, however. A full experimental study will, of course, also have to include $b\bar{b}c\bar{c}$, $c\bar{c}c\bar{c}$, $b\bar{b}jj$, etc.

While such a search is challenging, and is necessary to find NLSPs only in a limited region of parameter space, $\Br(\tilde \chi^0_1 \to h + \tilde G)$ is generically order-one whenever $m_{\tilde \chi^0_1}$ is sufficiently far above $m_h$. Thus, this search is worth pursuing for the prospect of finding the Higgs, especially if other channels like $Z(\ell^+ \ell^-)+\met$ begin to show an excess. Finding decays to Higgs would also be an important step in establishing the Higgsino-like nature of any prospective NLSP candidate.

\section{Conclusions}\label{sec:conclude}
\setcounter{equation}{0} 

\subsection{Summary of Results}

In this paper, we have analyzed in detail the signatures of promptly-decaying, general neutralino NLSPs at the Tevatron. Through both simple $S/\sqrt{B}$ considerations and detailed simulations, we have derived the most promising Tevatron search channels for each type of general neutralino NLSP: $\gamma\gamma+\met$, for bino or very light Higgsino NLSP; $W(\ell\nu)+\gamma+\met$, for wino co-NLSPs; $Z(\ell\ell)+\met+X$ and possibly $Z(\ell\ell)+Z(\ell'\ell')+\met$ for $Z$-rich Higgsinos; and  possibly multi-$b$+$\met$ for Higgs-rich Higgsinos. 

The estimated limits from existing Tevatron searches, along with the projected limits in 10 fb$^{-1}$ for optimized searches, are presented in Table \ref{tab:summary} for the cases of pure bino, wino, and Higgsino NLSPs. More generally, there are limits on mixed  NLSPs arising from $\gamma\gamma+\met$ which cannot be easily summarized in the table, but are displayed in Figure \ref{fig:exclggmetgen}. Note that for bino NLSPs, the result is expressed as a limit on the lightest chargino mass, while in the other cases it is a limit on $m_{NLSP}$. However, in these other cases, $m_{NLSP} \approx m_{\tilde{\chi}^\pm_1}$, so we can always think of the limit as being on the lightest chargino mass. 

\begin{ctable}[
caption = {Summary of Results. All numbers are in GeV, and unless otherwise specified, refer to the NLSP mass.},
label = tab:summary,
pos = t]{|c|c|c|c|c|c|}
{
}
{
\hline
\small NLSP Scenario & \small Search Channel & \small Curr. Lim. & \small Proj. Lim. & \small Proj. 3$\sigma$\\ 
\hline
\hline
bino & $\gamma\gamma+\met$ & $m_{\tilde{\chi}^\pm_1} >$ 270  & $m_{\tilde{\chi}^\pm_1} >$300 & None\\
wino co-NLSP & $W(\ell \nu)+\gamma+\met$ & 135 & 170 & 160\\
$Z$-rich Higgsino & $Z(\ell^+ \ell^-)+\met+X$  & None &  170 & 160\\
& $Z(\ell^+ \ell^-)Z(\ell'^+\ell'^-)+\met$  & None & 160 (?) & 140 (?)\\
Higgs-rich Higgsino & multi-$b$+$\met$ & None & 160 (?) & (?)\\
\hline
}
\end{ctable}

Somewhat surprisingly, we also saw that the Tevatron still has an opportunity for 5$\sigma$ discovery of wino and Higgsino NLSPs. We found that in the optimized $W(\ell\nu)+\gamma+\met$ and $Z(\ell^+\ell^-)+H_T+\met$ channels (and perhaps in the very clean  $Z(\to \ell^+ \ell^-) Z(\to \ell'^+ \ell'^-)+\met$ channel), discovery is possible for wino and Higgsino NLSPs, respectively, up to $m_{NLSP}\approx 135-140$ GeV.

Mostly, the role of the Tevatron will be to exclude regions of parameter space, or more optimistically to build up a sample of interesting statistically significant events that falls short of discovery. Because an NLSP that is a good target for one search (e.g. $Z+\met+X$) has small branching fractions to other decay modes, the Tevatron will have limited ability to correlate interesting excesses in more than one channel. On the other hand, the events in a given channel may contain hints about the underlying physics that can be exploited in future searches at the LHC. For instance, as we have discussed, a $\gamma\gamma+\met$ excess with only soft jets in the rest of the event is suggestive of a light Higgsino NLSP, which will in general have a nonnegligible branching fraction to $Z+\tilde{G}$. At the Tevatron we find that $\gamma\gamma+\met$ and $Z+\met+X$ have little overlap in their reach, but perhaps the LHC can fare better in substantiating the Higgsino interpretation with searches involving $Z$ bosons, especially if it can produce SUSY events with a strong cross section.

\subsection{Future Directions}

In this paper we have focused on the specific scenario of prompt neutralino NLSPs at the Tevatron.  As discussed in the introduction, searching for this scenario is one of the most well-motivated ways that the Tevatron can look for gauge mediation.  However, there are other studies related to gauge mediation at the Tevatron, that could be interesting as well.  We have not exploited the possibility that additional supersymmetric states, such as sleptons or colored sparticles, could increase the reach for gauge mediation at the Tevatron.  The phenomenology of including these states in combination with a general neutralino NLSP might also reveal new blind spots in the Tevatron's search for supersymmetry.  For instance, a Higgs-rich Higgsino NLSP produced in the decay of a heavier colored sparticle may avoid standard constraints from jets + $\met$ searches.  There are also extensions of the general neutralino NLSP scenario that do not rely on additional sparticle states.  The most straightforward extension is to consider delayed neutralino NLSP decays rather than prompt.  In gauge mediation models this corresponds simply to raising the fundamental scale of supersymmetry breaking without having to change anything else.  Using the tracking capabilities at both Tevatron experiments, and the EM timing capabilities at CDF or EM pointing capabilities at \dz, the Tevatron can set interesting limits on delayed general neutralino NLSPs~\cite{llgmsb}.

While the focus of this paper has been the Tevatron, we don't intend to downplay the LHC's capabilities. It is natural to ask how well the LHC can do, and how quickly it will surpass the capabilities of the Tevatron. Given that the LHC has not yet collided beams, it may be quite some time before it will have delivered comparable amounts of data. 

Existing LHC studies are largely focused on the minimal gauge mediation scenario (though see Ref. \citep{BMTWLHC}). For instance, a recent ATLAS study~\citep{ATLASGMSB} of $\gamma\gamma+\met$ shows that this can be a very good search channel with even small amounts of luminosity. Recently ATLAS has considered the case of Higgsino NLSPs in the $Z(\to \ell^+ \ell^-) \gamma$ channel \citep{Harper:2009iw}, estimating a reach of about 135 GeV with 3 fb$^{-1}$ at $\sqrt{s}=10$ TeV. We have seen that the Tevatron can already exclude this region of parameter space.

The main advantage of the LHC over the Tevatron would be the ability to make gluinos and squarks; if strongly-produced particles are accessible, there is a capacity for a much higher rate. However, we should not be too wedded to particular ideas about the SUSY spectrum. The gluinos and squarks could be sufficiently heavy that this is not a strong advantage\footnote{In both ATLAS studies cited in this paper the gluino and squark masses for the model points studied were $\mathcal{O}(1)$ TeV.}. For some channels, like searches involving $h \to b\bar{b}$, QCD backgrounds at the LHC can be more severe; the Tevatron might be a slightly cleaner environment (though, naturally, any channel in which QCD backgrounds are dominant will be an uphill struggle to achieve statistically significant results at any hadron collider).

The issues of backgrounds at the LHC and a detailed comparison of LHC and Tevatron reach deserve further study. We plan to examine these and other issues in detail in a forthcoming paper \citep{gmsblhc}. In particular, gauge mediation signatures at the LHC along the lines of the $Z+\met+X$ and multi-$b$+$\met$ signatures that we discussed in this paper have received little attention so far. It would be interesting to consider not only simple extrapolation of the analyses in this paper to the LHC, but also more sophisticated searches using substructure techniques along the lines of \citep{Butterworth:2008iy} to look for boosted $Z$ or $h$ bosons. Also, the detectors have timing and pointing capabilities that should be exploited to their fullest to analyze the case of non-prompt NLSP decays \citep{llgmsb}.

\section*{Acknowledgments}
We thank Yuri Gershtein, John Hobbs, Eunsin Lee, Seog Oh, Sasha Pronko, Dave Toback, Neal Weiner and Junjie Zhu for helpful discussions, and Andrey Katz, Sasha Paramonov, Scott Thomas and Brock Tweedie for discussions and comments on the manuscript. We also acknowledge the Galileo Galilei  Institute for Theoretical Physics in Florence, Italy, where part of this work was completed. The work of PM was supported in part by NSF grant PHY-0653354 and DOE grant  DE-FG02-90ER40542.   The work of DS was supported in part by DOE grant DE-FG02-90ER40542 and the William D. Loughlin membership at the Institute for Advanced Study.

\bibliographystyle{utphys}
\bibliography{nlsp3}

\end{document}